\documentclass{aa}

\usepackage{graphicx}
\usepackage{natbib}
\usepackage{ulem}
\usepackage{lscape} 
\usepackage[varg]{txfonts}

\bibpunct{(}{)}{;}{a}{}{,} 

\usepackage{epsfig}

\usepackage{amssymb}

\usepackage{amsmath}
 
\usepackage{xcolor} 

\usepackage{xspace} 

\newcommand{\ind}[1]{_{\mathrm{#1}}}

\newcommand{\D}{\displaystyle}

\defcitealias{Balona_2015}{Bal15}
\defcitealias{Ceillier_2017}{Cei17}
\defcitealias{Costa_2015}{Cos15}
\defcitealias{Davenport_2016}{Dav16}
\defcitealias{Deacon_2016}{Dea16}
\defcitealias{Godoy-Rivera_2018}{GR18}
\defcitealias{Howell_2016}{Hwl16}
\defcitealias{Honda_2016}{Hnd16}
\defcitealias{Katsova_2018}{Kat18} 
\defcitealias{Kumar_2018}{Kum18} 
\defcitealias{Pugh_2016}{Pug16}
\defcitealias{Schwamb_2012}{Sch12}
\defcitealias{Tang_2012}{Tan12}
\defcitealias{Tayar_2015}{Tay15}

\begin{document}

\title{Active red giants: close binaries versus single rapid rotators}


\author{Patrick Gaulme\inst{\ref{inst1},\ref{inst2}} 
\and Jason Jackiewicz\inst{\ref{inst2}} 
\and Federico Spada \inst{\ref{inst1}} 
\and Drew Chojnowski\inst{\ref{inst2}} 
\and Beno\^{\i}t Mosser \inst{\ref{inst6}} 
\and Jean McKeever\inst{\ref{inst3}} 
\and Anne Hedlund\inst{\ref{inst2}} 
\and Mathieu Vrard\inst{\ref{inst5},\ref{inst6}} 
\and Mansour Benbakoura\inst{\ref{inst7},\ref{inst8}} 
\and Cilia Damiani\inst{\ref{inst1}}
}

\institute{Max-Planck-Institut f\"{u}r Sonnensystemforschung, Justus-von-Liebig-Weg 3, 37077, G\"{o}ttingen, Germany \email{gaulme@mps.mpg.de}\label{inst1}
\and 
Department of Astronomy, New Mexico State University, P.O. Box 30001, MSC 4500, Las Cruces, NM 88003-8001, USA \label{inst2}
\and
Department of Astronomy, Yale University, P.O. Box 208101, New Haven, CT 06520-8101, USA \label{inst3}
\and
LESIA, Observatoire de Paris, Universit\'e PSL, CNRS, Sorbonne Universit\'e, Universit\'e de Paris, 92195 Meudon, France\label{inst4}
\and
Department of Astronomy, The Ohio State University, 140 West 18th Avenue, Columbus OH 43210, USA \label{inst5}
\and
Instituto de Astrof\'isica e Ci\^encias do Espa\c{c}o, Universidade do Porto, CAUP, Rua das Estrelas, 4150-762, Porto, Portugal\label{inst6}
\and
IRFU, CEA, Universit\'e Paris-Saclay, F-91191 Gif-sur-Yvette Cedex, France\label{inst7}
\and
Universit\'e Paris Diderot, AIM, Sorbonne Paris Cit\'e, CEA, CNRS, F-91191 Gif-sur-Yvette Cedex, France\label{inst8}
}

\titlerunning{Active red giants: close binaries versus single rapid rotators}
\authorrunning {Gaulme et al.}
\abstract{
Oscillating red-giant stars have provided a wealth of asteroseismic information regarding their interiors and evolutionary states, and access to their fundamental properties enable detailed studies of the Milky Way. The objective of this work is to determine what fraction of red-giant stars shows photometric rotational modulation, and understand its origin. One of the underlying questions is the role of close binarity in this population, standing upon the fact that red giants in short-period binary systems (less than 150 days or so) have been observed to display strong rotational modulation. 
We select a sample of about 4500 relatively bright red giants observed by Kepler, and show that about 370 of them ($\sim 8\%$) display rotational modulation. Almost all have oscillation amplitudes below the median of the sample, while 30 of them are not oscillating at all. Of the 85 of these red giants with rotational modulation chosen for follow-up radial-velocity observation and analysis, 34 show clear evidence of spectroscopic binarity. Surprisingly, 26 of the 30 non-oscillators are in this group of binaries. To the contrary, about 85\,\% of the active red giants with detectable oscillations are not part of close binaries. With the help of stellar masses and evolutionary states computed from the oscillation properties, we shed light on the origin of their activity. It appears that low-mass red-giant branch stars tend to be magnetically inactive, while intermediate-mass ones tend to be highly active. The opposite trends are true for helium-core burning (red clump) stars, whereby the lower-mass clump stars are comparatively more active and the higher-mass ones less so. In other words, we find that low-mass red-giant branch stars gain angular momentum as they evolve to clump stars, while higher-mass ones lose angular momentum. The trend observed with low-mass stars leads to possible scenarios of planet engulfment or other merging events during the shell-burning phase. Regarding intermediate-mass stars, the rotation periods that we measure are long with respect to theoretical expectations reported in the literature, which reinforces the existence of an unidentified sink of angular momentum after the main sequence. This article establishes strong links between rotational modulation, tidal interactions, (surface) magnetic fields, and oscillation suppression.  There is a wealth of physics to be studied in these targets not available in the Sun.
}
\keywords{(Stars:) binaries: spectroscopic - Stars: rotation - (Stars:) starspots - Stars: oscillations - Techniques: spectroscopy - Techniques: radial velocities - Techniques: photometric - Methods: observational - Methods: data analysis }

\maketitle


\section{Introduction}
\label{sect_intro}
Thanks to the observations of the space photometers CoRoT \citep{Baglin_2009}, Kepler \citep{Borucki_2011}, and now TESS \citep{Ricker_2015}, red giant (RG) stars are known to commonly display solar-like (SL) oscillations \citep{DeRidder_2009}. As for main-sequence (MS) SL stars, these global oscillations are stochastically excited by turbulent convection in their outer convective envelopes. From a revised classification of the Kepler targets \citep{Berger_2018} based on the second data release of the ESA GAIA mission \citep{Gaia_Collaboration_2018} and using the revised temperatures from the Kepler Stellar Properties Catalog \citep[KSPC DR25,][]{Mathur_2017}, \citet{Hon_2019} detected oscillations in about 92\,\% of the $21,000$ RGs observed by the original Kepler mission. We note that this fraction is much larger than the 40\,\% detection rate reported for MS SL stars \citep{Chaplin_2011a,Mathur_2019}, whose light curves are dominated by observational noise \citep{Mosser_2019}. Asteroseismology is thus a very powerful tool to analyze the evolution of stars after the MS as it can be performed on almost all targets of vast samples of stars.

Contrarily to MS SL stars \citep[e.g.,][]{Garcia_2014b}, only a few percent of the RGs show a pseudo-periodic photometric variability likely originating from surface spots \citep{Ceillier_2017}, which we hereafter call ``rotational modulation''. The most commonly adopted explanation is that RGs are slow rotators \citep[e.g.,][]{Gray_1981,de_Medeiros_1996} where no significant dynamo-driven magnetic fields are generated in their large convective envelope, implying no detectable surface spots. In this respect, it is worth noticing that no direct detection of magnetic field has been reported so far for stars on the red giant branch (RGB) with masses below 1.5 $M_\odot$ \citep{Charbonnel_2017}.
Nevertheless, measuring the rotation rate of RG stars is of prime importance to understand the evolution of stellar rotation after the MS. For this reason, much effort has lately been devoted to measure their rotation rates from their radiative cores \citep[e.g.,][]{Beck_2012, Mosser_2012a,Deheuvels_2015,Gehan_2018} to their surfaces \citep[e.g.,][]{Carlberg_2011,Costa_2015,Deheuvels_2015,Tayar_2015,Ceillier_2017}.

From high-resolution (HR) spectroscopic measurements of projected rotational velocities $v\sin i$ by \citet{Carlberg_2011}, about 2\,\% of  RGs are rapid rotators, using a threshold of $v\sin i=10$ km s$^{-1}$ as is commonly adopted in the literature \citep[e.g.,][]{Drake_2002,Massarotti_2008,Costa_2015,Tayar_2015}. Although the details of stellar dynamo theory are not yet fully understood, the empirical correlation between magnetic activity and rotation \citep[e.g.,][]{Noyes_1984} lends support to the expectation that the fraction of stars with detected rotational modulation should be similar to those with rapid rotation \citep[see also][]{Ceillier_2017}.
Despite this correlation, as of yet there is no theoretical calculation that finds the condition for spot generation to be $v\sin i \geq 10$ km s$^{-1}$. However, \citet{Ceillier_2017} reported the detection of starspot variability in about 2\,\% of the light curves from a systematic search in the Kepler oscillating RGs, which matches the fraction of rapid RG rotators of \citet{Carlberg_2011}. In the rest of the paper, we consider that RGs with rotational modulation are fast rotators relatively to the bulk of RG stars.

Currently, the observed rotation in both intermediate and low mass stars present puzzling features compared with the standard theoretical predictions. Intermediate mass stars ($M \in [1.5-3]\,M_\odot$) do not have a convective envelope during the MS, which means no large scale magnetic fields, and hence no angular momentum loss during it \citep[e.g.,][]{Durney_Latour_1978,Tayar_2015}. The rotation periods of A and late B-type stars cover a very broad range during the MS up to $v\sin i \sim 300$ km s$^{-1}$ \citep{Zorec_Royer_2012}. When the rapid rotators reach the RGB, they develop a convective envelope while still spinning fast, and they keep having a significantly large rotation ($v \sin i\geq 10$ km s$^{-1}$) during the subsequent core-helium-burning stage, also known as the red clump (RC) in reference to the color-magnitude diagram \citep[e.g.,][]{Girardi_2016}. This theoretical expectation is actually not observed in practice: \citet{Ceillier_2017} find 1.9\,\% of $M\geq2M_\odot$ RGs with rotational modulation, and \citet{Tayar_Pinsonneault_2018} report a distribution of $v\sin i$ peaking between 3 and 5 km s$^{-1}$ with maximum values at 10 km s$^{-1}$. 
As regards low-mass stars ($M\leq1.3 M_\odot$), which have convective envelopes and thus lose angular momentum during the MS, we expect slow rotation rates when they evolve. From stellar evolution models, \citet{Tayar_2015} indicate that low-mass stars should show $v\sin i \leq 0.3$ km s$^{-1}$ during the RGB, and at maximum an order of magnitude larger on the RC. Surprisingly, measurements of $v\sin i$ by \citet{Tayar_2015} and photometric modulation by \citet{Ceillier_2017} reported that between 7\,\% and 15\,\% of RGs with masses $M\leq1.1 M_\odot$ are fast rotators. Such low-mass rapidly rotating RGs are thus suspected to have either been spun up by tidal interactions with a stellar companion, or have merged with or engulfed a stellar or sub-stellar companion.

In parallel to this, it has also been observed that RGs in close eclipsing binary (EB) systems have peculiar photometric properties \citep{Gaulme_2014}. Among the 35 Kepler RGs that are confirmed to belong to EBs, 18 display regular solar-like oscillations -- with their amplitudes matching the empirical expectations \citep[e.g.,][]{Kallinger_2014} -- and no rotational modulation. All of the remaining systems display rotational modulation, 7 with partially suppressed oscillations, and 10 with no detectable oscillations \citep[][Benbakoura et al. submitted]{Gaulme_2014, Gaulme_2016a}. \citet{Gaulme_2016a} showed that the non-detection of oscillations is not an observational bias. 
This is observed in the closest systems, where most orbits are circularized, and rotation periods are either synchronized or in a spin-orbit resonance. Such a configuration is observed for systems whose orbital periods are shorter than about $P\ind{orb}\sim150$ days, and the sum of the stellar radii relatively to the system's semi-major axis $(R_1+R_2)/a \gtrsim 7\,\%$ of the semi-major axis. In the following, we consider a binary system to be close when $P\ind{orb}\lesssim200$ days. 
\citet{Gaulme_2014} suggested that the surface activity and its concomitant mode suppression originate from tidal interactions. During the RGB, close binary systems reach a tidal equilibrium where stars are synchronized and orbits circularized \citep[e.g.,][]{Verbunt_Phinney_1995, Beck_2018}. Red giant stars 
are spun up during synchronization, which leads to the development of  a dynamo mechanism inside the convective envelope, leading to surface spots. The magnetic field in the envelope likely reduces the turbulent excitation of pressure waves by partially inhibiting convection. Since spots can also absorb acoustic energy, these two effects lead to the suppression of oscillations.

This phenomenon observed with RGs in EBs 
led us to hypothesize
that a large fraction of the RGs with rotational modulation -- hereafter ``active'' RGs -- and suppressed oscillations actually belong to short-period (non-eclipsing) binary systems. This connection  between fast rotating RGs and binarity has been considered for a long time \citep[e.g.,][]{Simon_Drake_1989,Carlberg_2011}, but no dedicated observations have been led to test it. Only archived data from large surveys were considered in recent works \citep{Tayar_2015,Ceillier_2017}, leading to rather inconclusive results on that matter. 

The goal of the present paper is firstly to determine what fraction of active RGs are in close binary systems, and secondly to understand the origin of magnetic activity in the RGs that do not belong to close binaries. We base our work on a self-consistent approach where we select a sample of about 4500 RGs observed by Kepler that are not known to belong to any binary system, and that are expected to exhibit SL oscillations 
 (Sect. \ref{sect_2}). 
Among them, we show that about 370 ($\approx8\,\%$) display significant rotational modulation, which appears to be correlated with weak oscillations (Sect. \ref{sec_surf_mod}). We then look for radial velocity (RV) variability in a subset of 85 targets from HR spectroscopic measurements (Sect. \ref{sec_SB_frac}), and find a significant fraction of spectroscopic binaries (SBs).
In the last section (\ref{sect_disc}), we classify and quantify the possible scenarios that could have led to the active RGs that are not in binary systems.

\section{Data and methods}
\label{sect_2}
\subsection{A sample of 4500 RGs observed by Kepler}
\begin{figure}[t!]
\includegraphics[width=8.5cm]{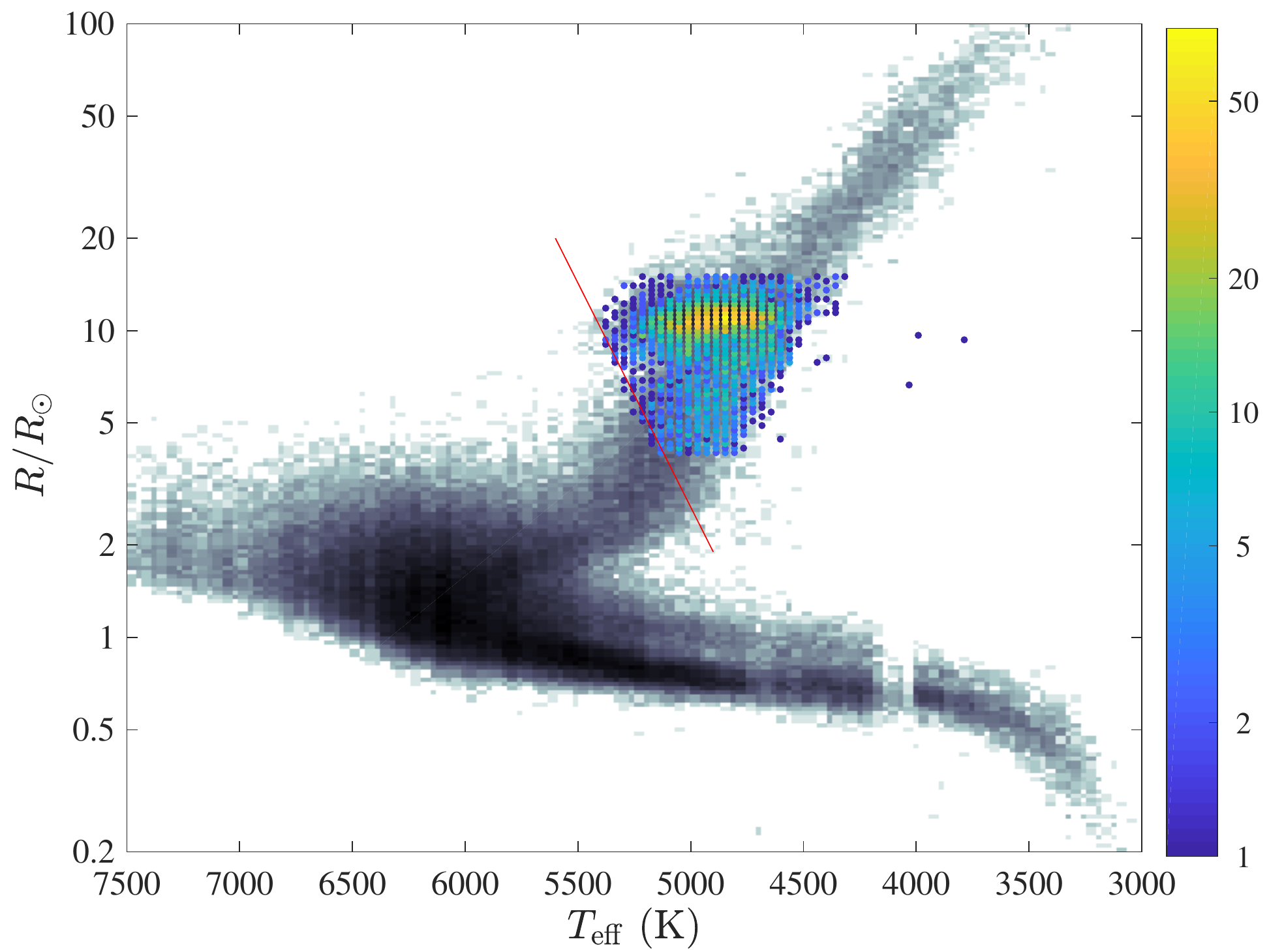} 
\caption{Radius-temperature diagram of the stars available in \citet{Berger_2018} in the range $[0.2,100]\,R_\odot$ and $[3000, 7500]$ K. Radii are from GAIA and temperatures from the Kepler MAST archive. Temperature axis is reverse, that is with higher $T\ind{eff}$ toward left. Color-coding represents logarithmic number density. The colored part of the diagram indicates the location of our sample. The red line indicates the limit between subgiant and giant stars according to \citet{Berger_2018}, such as the RGs lay at the top-right corner. According to \citet{Berger_2018}, the discontinuity in $T\ind{eff}$ near 4000 K is an artifact due to systematic shifts in $T\ind{eff}$ scales in the DR25 Kepler Stellar Properties Catalog.}
\label{fig_sample_selec}
\end{figure}

The first step of our work consists of measuring the fraction of RGs displaying clear rotational modulation, and reassessing the correlation between rotational modulation and suppression of SL oscillations pointed out by e.g., \citet[][]{Garcia_2010}, \citet[][]{Chaplin_2011a}, \citet[][]{Gaulme_2014}, or \citet{Mathur_2019}. Our main concern at this step is to avoid observational biases by including targets that are not suitable for our study. We started from the updated RG catalog of \citet{Berger_2018}, and selected a subsample of targets whose oscillations should be detectable provided that they are regular RGs. We base our selection criterion on the conclusions of \citet{Mosser_2019}, who quantified the asteroseismic performance of Kepler as a function of apparent magnitude and observation duration. This led us to consider RGs with Kepler magnitude $m\ind{Kep} \leq 12.5$ and observation duration longer than 13 quarters ($\approx 3$ years). We also limit ourselves to RGs with radii less than $15 R_\odot$ to avoid being biased by targets whose oscillations at maximum amplitude are at frequencies $\nu\ind{max} \lesssim15\ \mu$Hz, which could be filtered out during light-curve processing. We also exclude those with radii less than $4 R_\odot$ to avoid oscillation modes beyond Kepler's long-cadence Nyquist frequency ($\approx 283\,\mu$Hz). We note that 75\,\% of the Kepler RG sample have radii between 4 and 15 $R_\odot$. In other words we focus on the ``mainstream'' RGs. A total of 4580 out of the $\approx 21,000$ RGs meet these criteria (See Fig. \ref{fig_sample_selec}). 

As misclassification can happen in large catalogs, we excluded stars that were clearly not single RGs from a visual inspection of the Kepler light curves and their frequency spectra. This downselection filtered out obvious binary systems as double RG oscillators, eclipsing binaries, ellipsoidal binaries, and highly eccentric non-eclipsing binaries \citep[``heartbeat'' stars'', e.g., ][]{Welsh_2011,Beck_2014}, but also classical pulsators as $\delta$ Scuti or $\gamma$ Doradus.  We found 88 light curves including a binary signal, and 21 with either a $\delta$ Scuti or a $\gamma$ Doradus pulsator. We identified another 6 with a damaged Kepler light curve, that is where photometry is clearly wrong. 
We thus ended up with a sample of 4465 stars classified as RGs according to \citet{Berger_2018} and with no obvious indication of misclassification or peculiarity. 

\begin{figure}[t!]
\includegraphics[width=9cm]{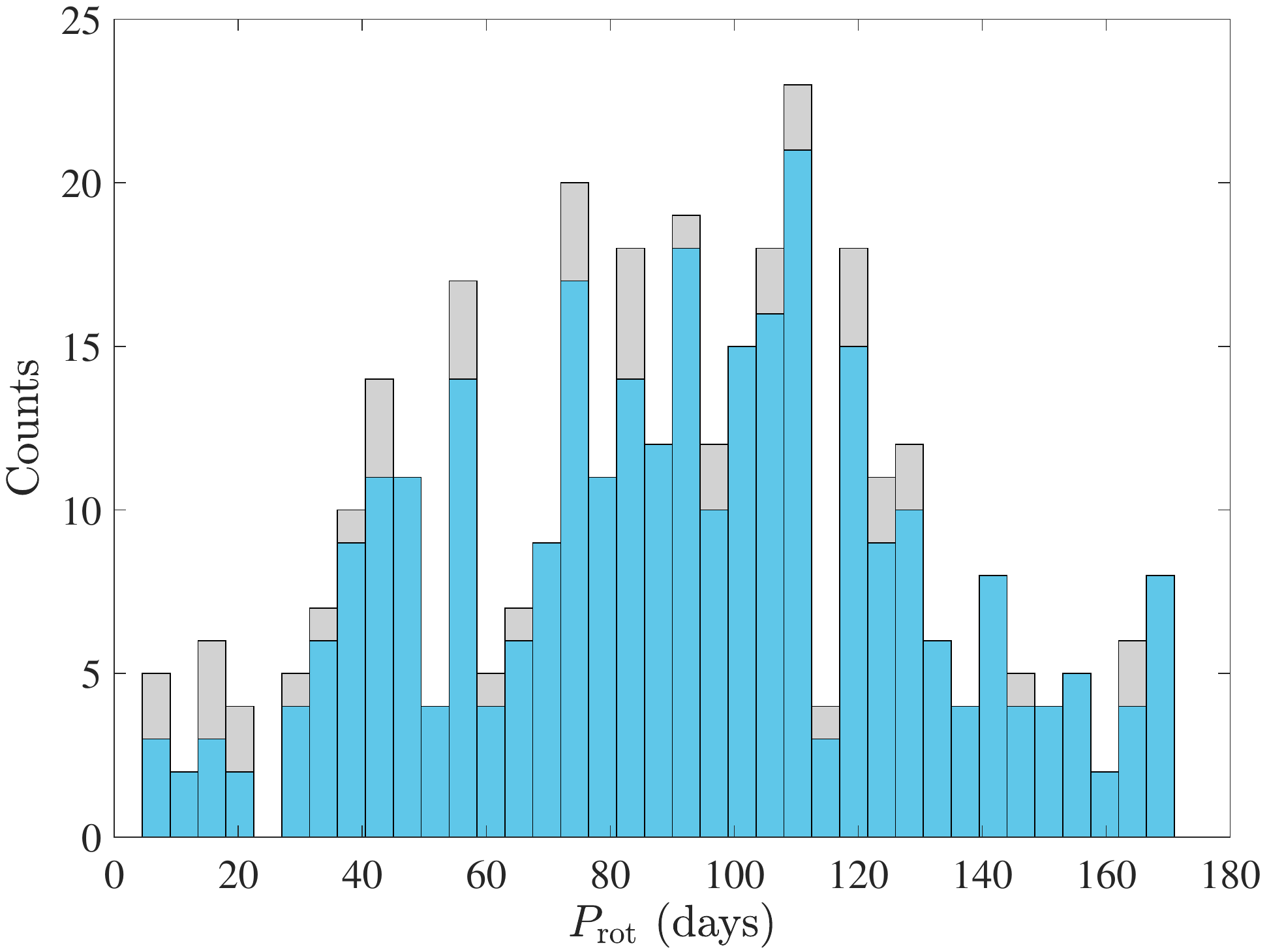} 
\caption{Histogram of the rotation periods $P\ind{rot}$ that we measure for 340 RGs where both oscillations and rotation are detected. The gray histogram includes all targets, whatever their crowding factor. The blue histogram includes only the 305 targets with crowding larger than 0.98. This figure echoes Fig. 5 of \citet{Ceillier_2017}.}
\label{fig_hist_Prot}
\end{figure}

\subsection{The Kepler light curves}
\label{sec_data_proc}

We primarily worked with the Kepler public light curves that are available on the Mikulski Archive for Space Telescopes (MAST)\footnote{\url{http://archive.stsci.edu/kepler/}}. Two types of time series are available: the Simple Aperture Photometry (SAP) and the Pre-search Data Conditioning Simple Aperture Photometry (PDC-SAP) light curves. The latter consist of time series that were corrected for discontinuities, systematic errors and excess flux due to aperture crowding \citep{Twicken_2010}. They do not meet our requirements for monitoring the rotational modulation, which is often removed during the process \citep[e.g.,][]{Garcia_2014,Gaulme_2014}. We thus made use of the SAP data to preserve any possible long-term signal. This choice entails our own detrending and stitching operation on the light curves while ensuring that the rotational modulation is preserved after each interruption of the time series. The methods employed to clean the time series are detailed in \citet{Gaulme_2016a}.

To reinforce our results -- especially the flagging of rotational modulation and the measurement of rotational periods --, we repeated the whole light processing and analysis with the public Kepler Light Curves Optimized For Asteroseismology \citep[KEPSEISMIC,][]{Garcia_2011,Pires_2015}\footnote{\url{https://archive.stsci.edu/prepds/kepseismic/}}, which are optimized for asteroseismic analysis. KEPSEISMIC light curves were available for 4444 out of the 4465 targets. Except for a few cases, the results obtained with either light curves were in agreement. In the following, our results reflect the final analysis done by working both the SAP and KESPSEISMIC time series.

\subsection{Surface rotation periods}
\label{sec_methods_rotation}
In a first step, the presence of rotational modulation is automatically searched, before being checked by visual inspection. Rotational modulation is considered whenever the standard deviation of the photometric time series is larger than 0.1\,\%. We note that this definition is equivalent to the photometric index $S\ind{ph}$ defined by \citet{Mathur_2014} in the case no photometric period is measured.
Then from both the power spectrum and the autocorrelation of the times series
, the algorithm automatically checks for surface activity, and provides an estimate of its fundamental period in case of positive detection. Rotational modulation is considered for periods ranging from 1 to 180 days. From the power spectrum, peaks are considered to be significant when the statistical null hypothesis test is less than 1\,\% with respect to the average background noise level. Autocorrelation is used to check the estimate obtained from the power spectrum, by verifying that the highest peak lies in the same period range within a factor 1/2 or 2, as misestimates can happen. Visual inspection was then used to decide the actual rotation period in case of disagreement between power spectrum and autocorrelation periods.

Beyond measuring rotation periods, we need to be careful with possible photometric contamination from nearby stars. Indeed, it can happen that a considered RG does not have any spot but that a neighboring MS stars display spots and that the light curves leak into each other. \citet{Ceillier_2017} tracked this possible source of errors by making use of the crowding parameter available on the Kepler database at MAST. The crowding parameter is defined as the fraction of the flux coming from the considered target in the light curve. 
The lower the crowding parameters, the higher the chance of contamination. In their case, they considered all the 17,377 RGs with detected oscillations at the time, which included many faint targets. Their Fig. 5 shows the comparison between the histogram of rotational periods with and without targets where the crowding factor is less than 0.98. In their case, they found that the targets with low crowding represent $32\,\%$ of their active RG sample, and are a significant source of bias as they constitute a clear group of outlying $P\ind{rot}$ with values less than 30 days. In our case, Fig. \ref{fig_hist_Prot} shows that the stars with a crowding factor less than 0.98 (13\,\% of our active RGs) do not constitute a well defined group, so we can legitimately consider that targets with a lower crowding factor are not due to contamination in most cases, and that we are not affected by this problem. It is not surprising since we focus on purpose on the brightest part of the sample to avoid observational biases of that kind. 

\subsection{Oscillation analysis}
\label{sec_ppties_sample}
\begin{figure}[t!]
\includegraphics[width=9cm]{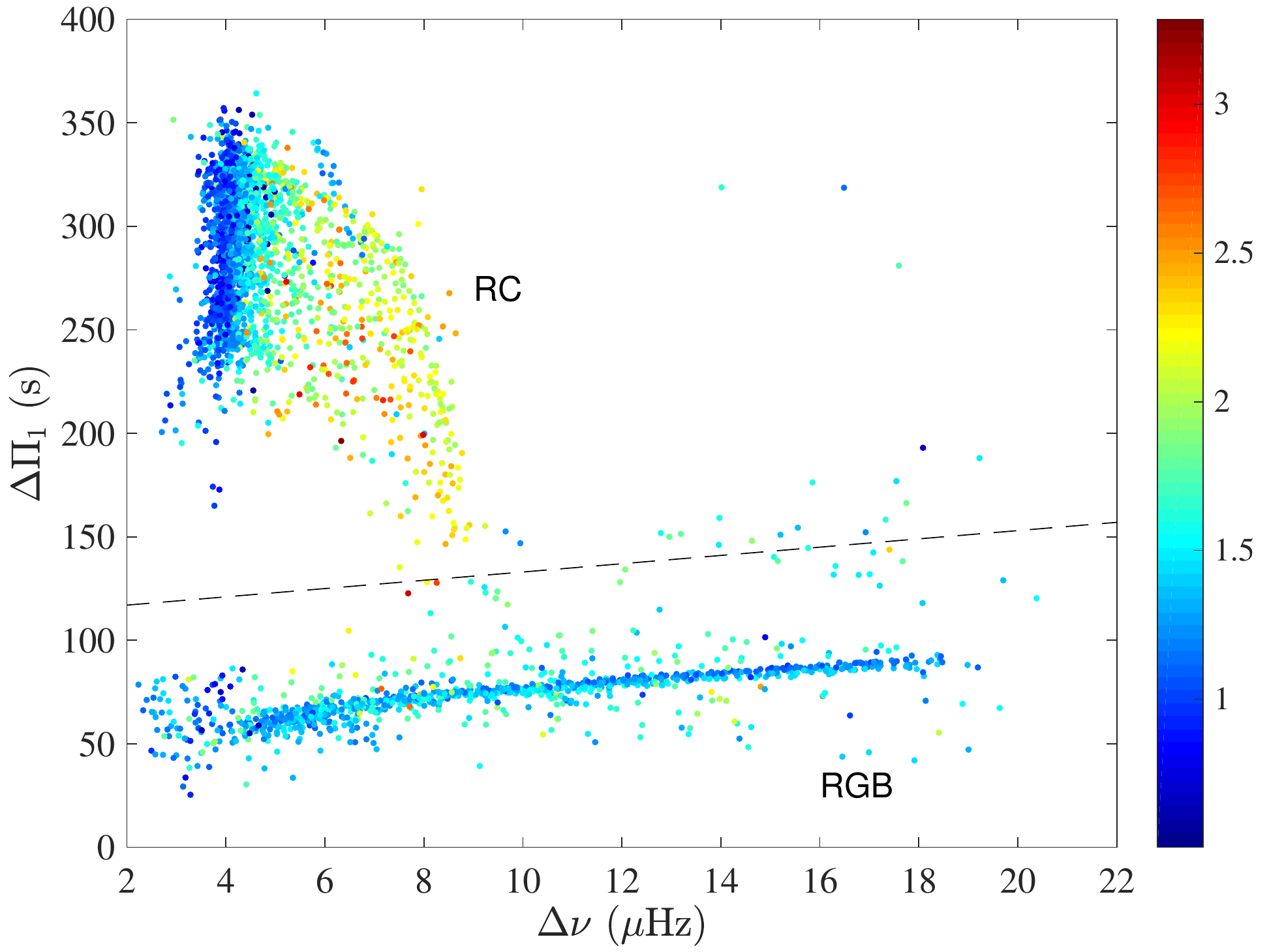} 
\caption{Period spacing of the $l=1$ mixed modes $\Delta\Pi_1$ (expressed in seconds) as a function of the large frequency separation $\Delta\nu$ (expressed in $\mu$Hz). The period spacing is measured for 3388 stars out of the sample of 4465 RGs. The colors of the markers indicate the stellar masses as indicated in the colorbar (deep blue is less that $1.5 M_\odot$). The dashed line separates the red clump stars (above) and the RGBs (below).}
\label{fig_Dpi_Dnu}
\end{figure}
\begin{figure}[t!]
\includegraphics[width=9cm]{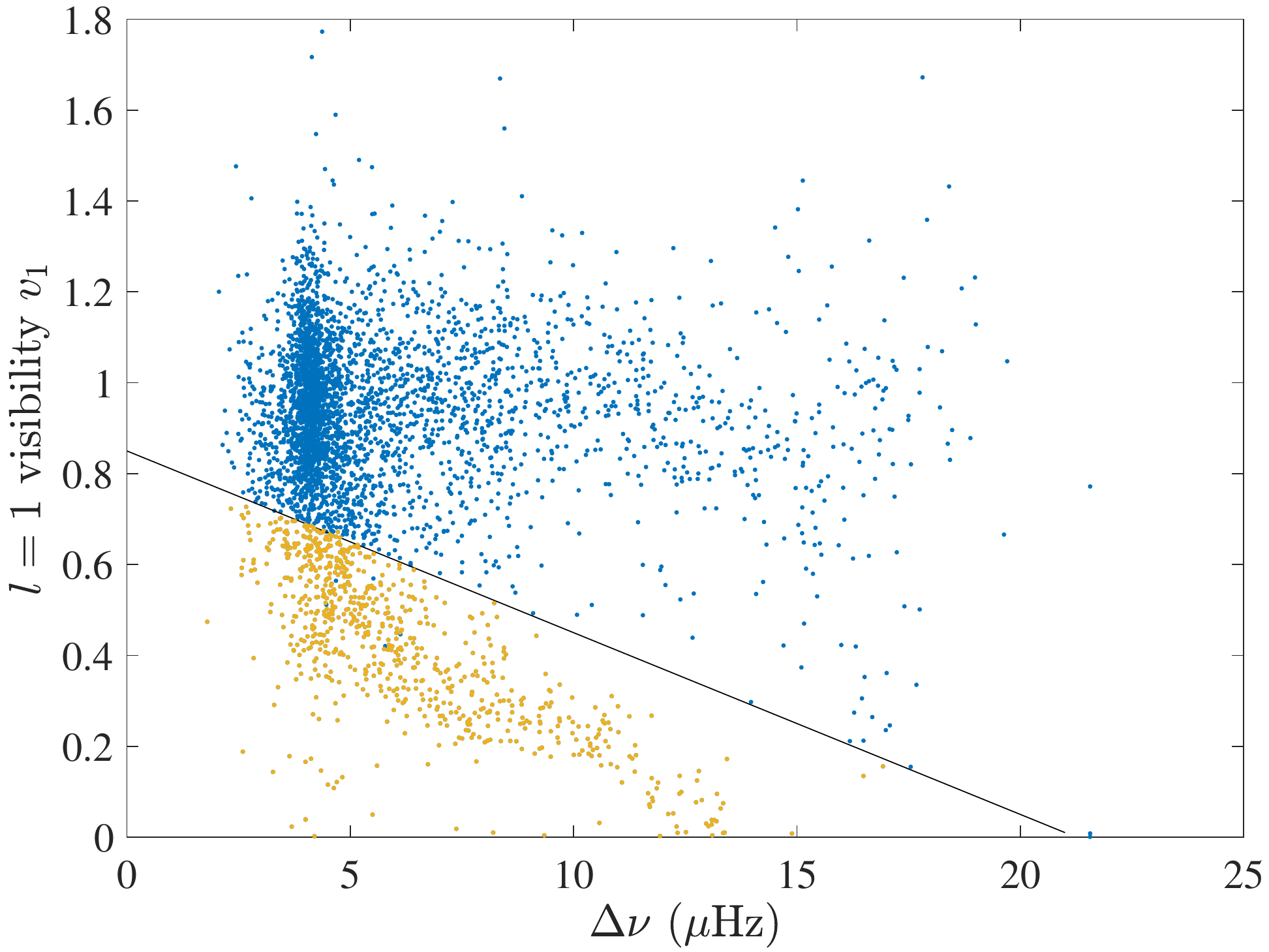} 
\caption{Visibility of dipole modes of the sample. Under the black line, $l=1$ modes are considered to be depleted.}
\label{fig_visibility}
\end{figure}
\begin{figure*}[ht!]
\center
\includegraphics[width=15.5cm]{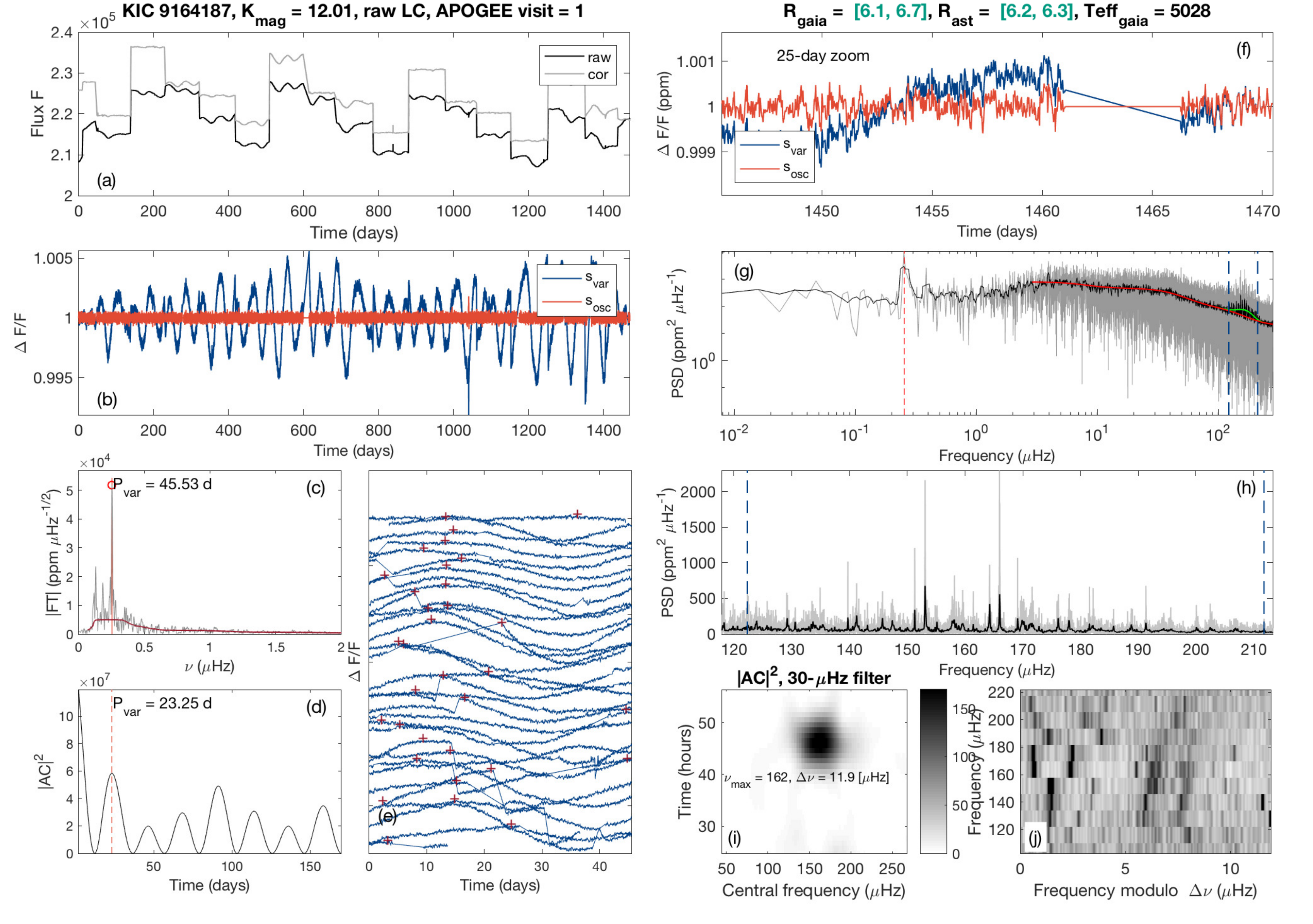} 
\caption{Data inspection tool applied to the RGB star KIC 9164187, which shows surface activity and oscillations, including mixed dipole modes. Left column from top to bottom. Panel (a): Kepler light curves as a function of time, where ``raw'' stands for SAP and ``cor'' for PDCSAP. Panel (b): light curves expressed in relative fluxes $\Delta F/F$, where the blue line contains the stellar activity and oscillations, and the red curve is optimized for oscillation search (activity filtered out).  Panel (c): Fourier spectrum of the time series, where the highest peak and its corresponding period is indicated in days. Panel (d): square of the autocorrelation $|AC|^2$ of the time series, where the highest correlation is indicated and expressed in days. 
Panel (e): light curve folded on the dominant variability period , where a vertical shift between consecutive multiples of the rotational period was introduced. Right column from top to bottom. Panel (f): zoom of the light curve over 25 days. Panel (g): log-log scale display of the power spectral density of the time series (gray line) expressed in ppm$^2\,\mu$Hz$^{-1}$ as a function of frequency ($\mu$Hz). The black line is a smooth of it (boxcar) over 100 points. The solar-like oscillations appear as an excess power between the two dashed blue lines. The plain red line represents the model of the stellar background noise and the green line is the Gaussian function used to model the oscillation power. The vertical dashed line indicates the peak corresponding to the rotational modulation determined in panel (c). Panel (h): power density spectrum of the time series as a function of frequency centered around $\nu\ind{max}$ (here $\approx162\ \mu$Hz). Panel (i): envelope of the autocorrelation function (EACF) as a function of frequency and time. The dark area corresponds to the correlated signal -- i.e., the oscillations -- where its abscissa indicates $\nu\ind{max}$ and its ordinate $2/\Delta\nu$. Panel (j): \'echelle diagram associated with the large frequency spacing automatically determined from the EACF plot. The power density spectrum is smoothed by a boxcar over three bins and cut into $\Delta\nu=11.9$-$\mu$Hz chunks; each is then stacked on top of its lower-frequency neighbor. This representation allows for visual identification of the modes. Darker regions correspond to larger peaks in power density. The $x$-axis is the frequency modulo the large frequency spacing (i.e., from 0 to $\Delta\nu$), and the $y$-axis is the frequency.}
\label{fig_DIT}
\end{figure*}

The detection of oscillations is based on the envelope of the autocorrelation function (EACF) developed by \citet[][]{Mosser_Appourchaux_2009}, which we apply to the power spectrum of the time series (Fig. \ref{fig_DIT} panel i). A first run of the EACF leads to initial estimates of the mean frequency separation $\Delta\nu$ and $\nu\ind{max}$, independently from the background noise level. The second step consists of refining $\Delta\nu$ and $\nu\ind{max}$ by taking into account the background noise.

The stellar granulation and accurate values of $\nu\ind{max}$ and mode amplitude $H\ind{max}$ are estimated by fitting the power spectrum as commonly performed in asteroseismology \citep{Kallinger_2014}, and already used in \citet{Gaulme_2016a}. Following \citet{Kallinger_2014}, the power density spectrum is fitted by the following function:
\begin{equation}
    S(\nu) = N(\nu) + \eta(\nu) \left[B(\nu) + G(\nu)\right],
\end{equation}
where the $N$ is the function describing the noise, $\eta$ is a damping factor originating from the data sampling, $B$ is the sum of three ``Harvey'' functions (super Lorentzian functions centered on 0), and G is the Gaussian function that accounts for the oscillation excess power:
\begin{equation}
    G(\nu) =H\ind{max} \exp{\left[-\frac{(\nu-\nu\ind{max})^2}{2\sigma^2}\right]}
\end{equation}
The terms $\nu\ind{max}$ and $H\ind{max}$ are the central frequency and height of the Gaussian function. 

A second estimate of $\Delta\nu$ is performed again with EACF from the whitened power spectral (power spectrum divided by background function). Oscillations are considered to be detected when both the maximum value of the EACF is larger than 8 \citep{Mosser_Appourchaux_2009} and peaks are visible in the spectrum.
We then compute the stellar masses and radii thanks to the asteroseismic scaling relations that were originally proposed by \citet{Kjeldsen_Bedding_1995} for SL MS oscillators, and then successfully applied to RGs \citep[e.g.,][]{Mosser_2013}. 
The stellar effective temperatures, also needed in the scaling relations, are those from the Kepler GAIA data release 2 with their corresponding errors. 
We employ the asteroseismic scaling relations as proposed by \citet{Mosser_2013} for RGs, i.e., where $\nu\ind{max,\odot} = 3104\ \mu$Hz, $\Delta\nu_\odot = 138.8\ \mu$Hz, $T\ind{eff,\odot} = 5777$ K, and where the observed $\Delta\nu$ is converted into an asymptotic one, such as $\Delta\nu\ind{as} = 1.038\,\Delta\nu$.


\begin{figure*}[t!]
\center
\includegraphics[width=14cm]{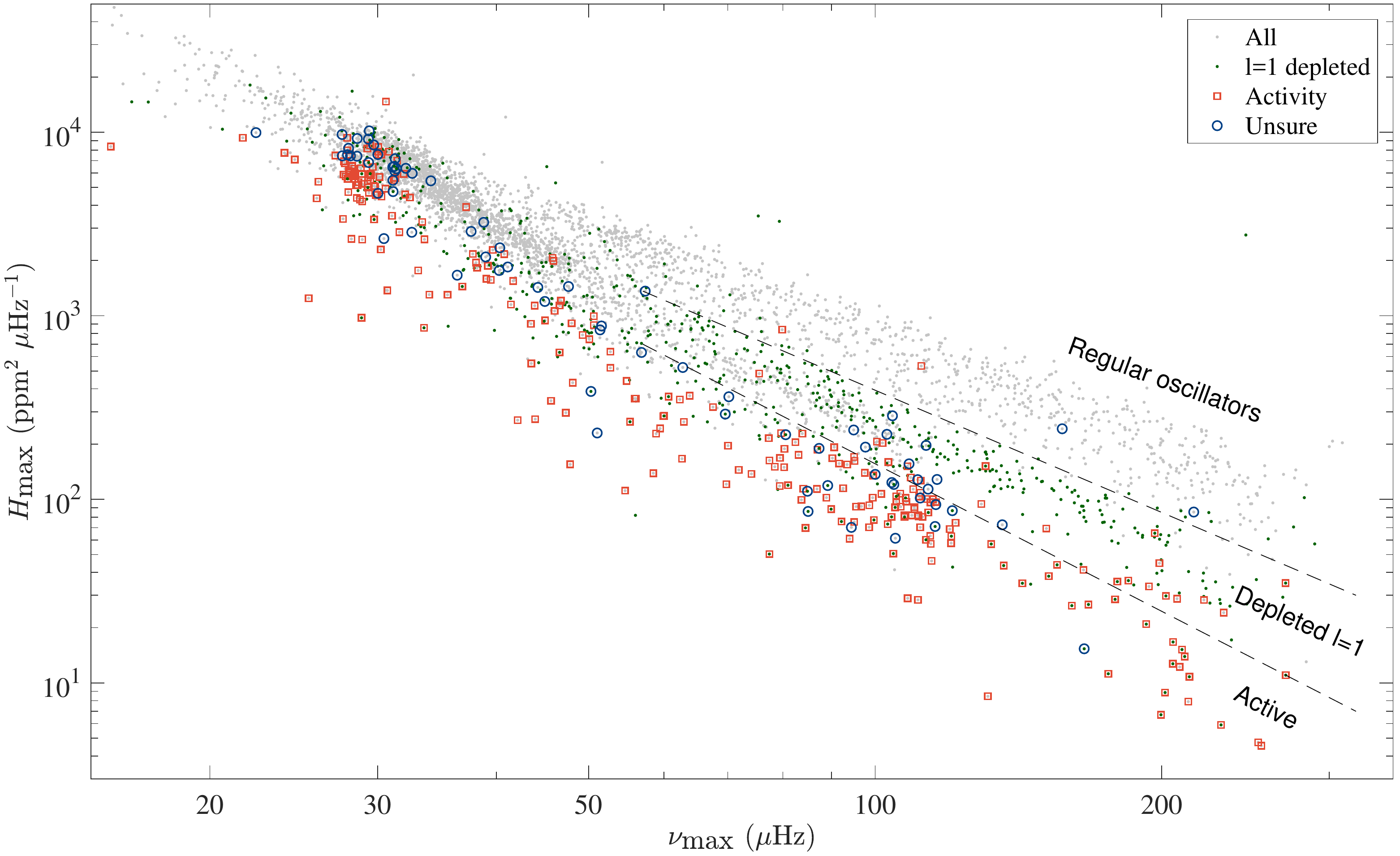}
\caption{Oscillation height $H\ind{max}$ (ppm$^2 \mu$Hz$^{-1}$) as a function of the frequency at maximum amplitude $\nu\ind{max}$ ($\mu$Hz) of the 4465 red giants considered in this work, minus the 30 for which no oscillations were detected, i.e. 4435 total. The parameter $H\ind{max}$ is the  height of the Gaussian function used to fit the oscillation contribution to the power spectral density. The frequency at maximum amplitude $\nu\ind{max}$ is the center of the same Gaussian function. 
Light gray points indicate all the targets, green are $l=1$ depleted, red squares are the systems with significant surface activity, and blue circles indicate the cases of ambiguous detection of surface activity. 
}
\label{fig_sample}
\end{figure*}

In addition to the properties of the pressure $p$ modes ($\nu\ind{max}$, $\Delta\nu$) we analyze the properties of the mixed modes, which result of the interaction between the $p$-modes that resonate in the convective envelope and the gravity $g$ modes that resonate in the radiative core. At first order, $p$ modes are evenly spaced by $\Delta\nu$ as a function of frequency, whereas $g$ modes are evenly spaced by $\Delta\Pi$ as a function of period. From the dipole ($l=1$) mixed modes it is possible to measure the mean period spacing of the dipole $g$ modes $\Delta\Pi_1$, which is related to the physical properties of the core. For RGs this information is precious as it allows us to unambiguously distinguish the H-shell (RGB) from the He-core (RC) burning stars. Since the CoRoT space mission, such a type of analysis is commonly performed with RG stars \citep[][]{Bedding_2011,Mosser_2011, Stello_2013}.
According to \citep{Mosser_2014} when $\Delta\nu \geq 9.5\ \mu$Hz, we assume the stars to be on the RGB. When mixed modes are detectable, the evolutionary status is determined according to the methods described in \citet{Vrard_2016}. Overall, we have the status of 3405 stars out of the 4465: 
we used 2307 published statuses that were already available in \citet{Vrard_2016}, and we determined the status of 1098 others. Figure \ref{fig_Dpi_Dnu} shows the period spacing of the dipole mixed modes $\Delta\Pi_1$ as a function of the large frequency separation $\Delta\nu$ for the subsample of active RGs. We note a larger dispersion of the period spacing $\Delta\Pi_1$ for RGBs with $\Delta\nu \leq 7\,\mu$Hz. This issue is well known \citep{Dupret_2009, Grosjean_2014}: RGBs with low $\Delta\nu$ are more evolved than those with larger values, and the coupling between $p$ and $g$ modes is less efficient. Hence, gravity-dominated mixed modes display lower signal-to-noise ratio. 

Finally, our analysis of the oscillation spectra also considers the question of the RGs with depleted dipole modes \citep[e.g.,][]{Mosser_2012b,Garcia_2014}, which is possibly caused by the presence of fossil magnetic fields in radiative cores of intermediate mass RGs that had convective cores during the MS \citep{Fuller_2015,Stello_2016,Mosser_2017}. Despite the suppression of dipole modes is a priori not related to rotational modulation -- the former is supposed to be caused by core magnetic fields, the latter by surface ones --, we thought it was interesting to flag it in case of an unlikely correlation. We applied the method presented by \citet[][]{Mosser_2012b} to measure the relative visibilities of the modes of degrees $l=0$, 1, 2, and 3 to the whole sample, whenever oscillations were detected.

\subsection{A dedicated pipeline to analyze the sample}
\label{sec_DIT}

We developed an automatic pipeline -- the Data Inspection Tool, hereafter DIT -- to analyze the whole sample, which cleans the light curves and extracts the properties of both rotational modulation and solar-like oscillations. The code is similar to that described in \citet{Gaulme_Guzik_2019} to look for stellar pulsators in eclipsing binaries. Beyond the measurements of rotational modulation, background and oscillations, the DIT produces a vetting sheet for each target that includes a series of figures allowing the user to appreciate the quality of the measurements. An example of vetting sheet is displayed in Fig. \ref{fig_DIT}.

All vetting sheets were visually inspected. The objective of the visual inspection was to double check the detection of rotational modulation, assess the value of the rotation period, and look for oscillations that had been missed by the DIT. Besides, it allowed us to exclude the obvious binaries and classical pulsators as explained in the previous section. Regarding rotational modulation, we did not strictly stick to the $S\ind{ph}\geq0.1\,\%$ threshold, as some photometric discontinuities could cause $S\ind{ph}$ larger than that. Besides, some actual photometric modulation could have lower $S\ind{ph}$. The distinction between a spurious photometric modulation, arising from imperfect light curve cleaning, and actual stellar signal was most of the time clear. However, in conditions of low signal-to-noise ratio (S/N), flagging the detection of stellar activity could be fragile. This is why we used two classes of flags for activity: clear and ambiguous detection (See Sect. \ref{sec_surf_mod}).

\subsection{High-resolution spectroscopy}
\label{sec_HR_spectro}
As indicated in the introduction, we study a subsample of RGs displaying rotational modulation with HR spectroscopy to look for RV variability. Our intend is not to perform a detailed analysis of these spectra such as retrieving chemical composition or $v \sin i$, which will be part of a future work. 

From October 2018 to June 2019, we were granted time on the \'echelle spectrometer of the 3.5-m telescope of the Astrophysical Research Consortium (ARC) at Apache Point observatory (APO), which covers the whole visible domain at an average resolution of $31,000$. This time allowed us to monitor a set of 51 RGs with clear rotational modulation.
Even though the ARC \'echelle spectrometer was not designed for fine RV measurements, it has successfully been used for this purpose in earlier works about RGs \citep[e.g.,][]{Rawls_2016,Gaulme_2016a}. The measurement error reported in these papers is about 0.5 km s$^{-1}$ for an RG spectrum with $S/N$ between 10 and 20. In practise our spectra have S/N ranging from 10 to 25. With ARCES data, we thus consider a target to be an SB when the dispersion is larger than $\sigma\ind{RV}=2$ km s$^{-1}$. The ARCES Optical spectra were processed and analyzed in the same way as in \citet{Gaulme_2016a} and we refer to this paper for details. 

Beyond the ARC telescope, we used infrared high-resolution spectroscopic data of another 34 targets that were available in the archived data \citep[release 16,][]{Ahumada_2019} of the APO Galactic Evolution Experiment \citep[APOGEE,][]{Majewski_2017}, which is part of the Sloan Digital Sky Survey IV \citep[e.g.,][]{Blanton_2017}. 
Regarding APOGEE data, we directly use the archived RVs. This spectrometer has an average resolution of $22,500$ and is known to be stable, with a noise level lower than 0.1 km s$^{-1}$ \citep{Deshpande_2013}. Knowing that the RV jitter of RGs with surface gravities $2\leq\log g\leq 3$  -- as we have here -- is less than 0.1 km s$^{-1}$ \citep{Hekker_2008}, we consider an RG to be an SB when the RV dispersion is larger than $\sigma\ind{RV} = 0.5$ km s$^{-1}$.


\section{Results}
\subsection{Occurrence of red giants with rotational modulation}
\label{sec_surf_mod}

\begin{figure}[t!]
\includegraphics[width=9cm]{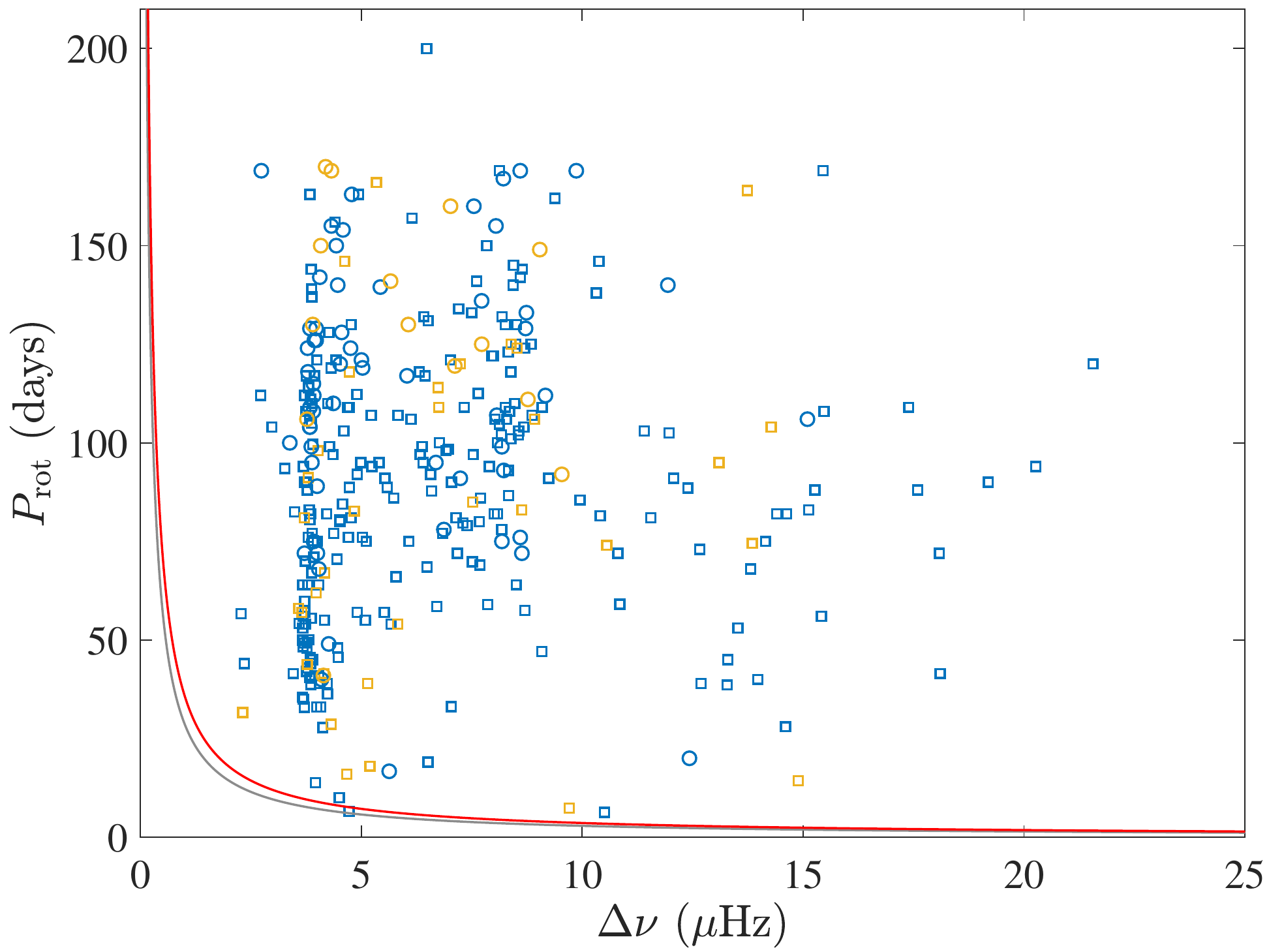} 
\caption{Rotation period $P\ind{rot}$ as a function of the mean large spacing $\Delta\nu$ for RGs displaying both rotational modulation and oscillations. The gray line indicates the critical period $T\ind{crit}$ and the red line a rotational velocity of 80\,\% of $T\ind{crit}$. Blue squares are the RGs with clear rotational modulation, blue circles those with low $S/N$ detection. Orange symbols indicate the same for those with crowding factor less than 0.98. This figure echoes Fig. 4 of \citet{Ceillier_2017}.}
\label{fig_Prot_Dnu}
\end{figure}

Among the 4465 RG stars that we have selected, our main conclusions regarding their Kepler light curves are the following:
\begin{enumerate}
\item A total of 298 RGs display clear rotational modulation, 
i.e., in 6.65\,\% of the cases. For another 73 targets (1.63\,\%), the detection of surface activity is ambiguous (low signal-to-noise ratio, $S/N$). If we assume the ambiguous detections as a proxy of our error in judgment, we can consider that $7.5\pm0.8$\,\% of the RGs display rotational modulation. We thus detect from three to four times more RGs with rotational modulation than the 2\,\% reported by \citet{Ceillier_2017}. As indicated in Sect. \ref{sec_methods_rotation}, the main difference with their work is that they considered the whole sample of oscillating RGs known at the time, which included faint stars and likely discarded stars with suppressed oscillations. By comparing Fig. \ref{fig_Prot_Dnu} with Fig. 4 of \citet{Ceillier_2017}, we clearly detect more stars in the $\Delta\nu$ range $[6.5, 9.5]\,\mu$Hz, which contains many intermediate-mass RCs. 
    \item  We detect oscillations in 99.32\,\% of our sample, i.e. in all except 30, which is much more than previously reported. This difference is not surprising because \citet{Hon_2019} considered the whole 21,000 RG catalog which contains faint stars, including crowded fields. Besides, their automatic pipeline may have encountered difficulties when dealing with low $S/N$ oscillation spectra, especially in the presence of rotational modulation. We interestingly note that the fraction 8\,\% of RGs where \citet{Hon_2019} do not detect oscillations is equal to the fraction of our RGs displaying activity. 
    \item All of the 30 targets with no detectable oscillations display a significant rotational modulation. In other words, if an RG does not oscillate, rotational modulation is always detected.  
    \item Almost all the active RGs display unusually low-amplitude oscillations, as well as most of those with ambiguous status. This can be seen in Fig. \ref{fig_sample}, which plots a proxy of the oscillation amplitude -- the height $H\ind{max}$ of the Gaussian function employed to fit the oscillation envelope in the power spectral density of the time series -- as a function of $\nu\ind{max}$. We distinguish three groups of oscillators: the regular, the dipole-depleted, and the active ones. The dipole-depleted RGs appears as a distinct group because if $l=1$ modes are absent, the oscillation power that we fit with a Gaussian function is reduced\footnote{Dipole-depleted RGs would not appear as a distinct group if our proxy of the oscillation amplitude were the amplitude of radial modes.}.
    \item Measurements of dipolar mode visibilities as performed in \citet{Mosser_2012b} and \citet{Mosser_2017} reveal that 17\,\% and 10\,\% of inactive and active RGs display $l=1$ depleted modes, respectively. To complement these results, a visual flagging of depleted dipole modes led us to identify 12\,\% and 20\,\% in the same two groups. We conclude that the fraction of depleted dipolar-mode oscillators is very similar in both samples of inactive and active RGs (Fig. \ref{fig_sample}), and cannot be considered as a significant source of biases in our analysis.
\end{enumerate}

\begin{figure}[t!]
\includegraphics[width=9cm]{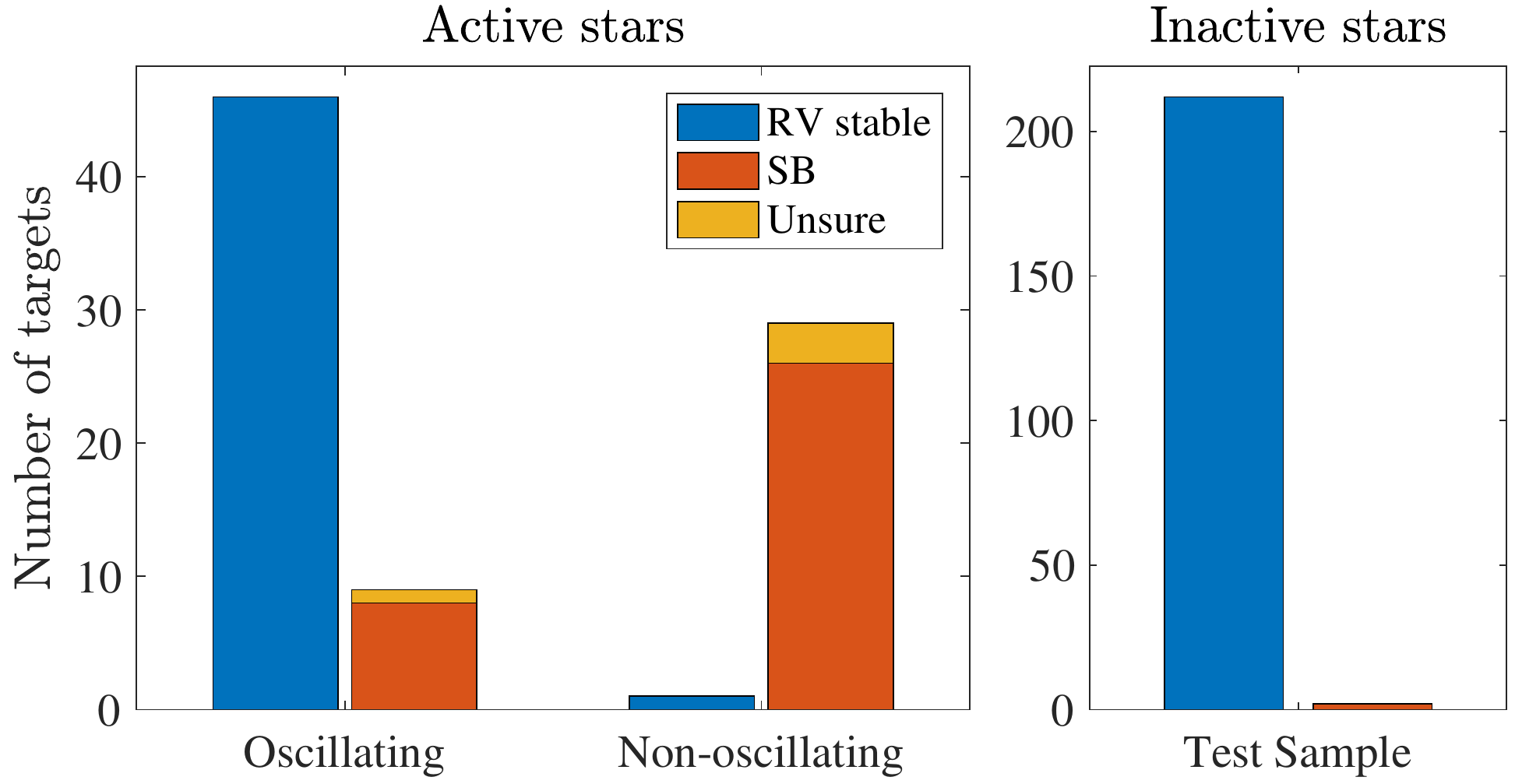} 
\caption{
Bar graphs showing how the spectroscopic targets divide into SBs and stars lacking in statistically significant RV variations (``RV stable''). Three groups are considered: the active stars with oscillations, the active stars without oscillations, and the inactive stars that all display clear oscillations. Among the 55 oscillating active RGs, 46 are RV stable, 8 clear SBs, and 1 unsure SB (orange). Among the 30 non-oscillating active RGs, 1 is RV stable, 26 are clear SBs and 3 are unsure SBs. Right panel is the same for the test sample of 212 RGs with no activity and oscillations that were observed more than four times by APOGEE. In the latter, 210 RGs are RV stable and 2 SBs.
}
\label{fig_frac_SB_active}
\end{figure}
\subsection{Occurrence of close binaries among the active red giants}
\label{sec_SB_frac}

According to the previous section, $7.5\pm0.8$\,\% of the RGs display rotational modulation. We now determine how many of them actually belong to close binary systems thanks to multi-epoch high-resolution spectroscopic measurements. 

The rate of SBs actually represents a lower limit of the fraction of close binaries because some systems are seen almost pole-on and display very low line-of-sight velocities. However, the RV semi-amplitude of the typical binary system that we expect -- synchronized, circularized, period of about 50 days, solar masses -- is larger than about 5 km s$^{-1}$ in 90\,\% of the cases, by assuming random inclination angles (see Appendix \ref{app_RV_theo}). 
In addition, we show that five measurements randomly distributed over a 200-day range with at least seven days between consecutive observations are very likely to detect RV variability. With the ARCES spectrometer, we observed four times most of our targets, and up to seven times some targets whose RV variability was confusing and needed more data. The number of visits per star is reported in Table \ref{table_ze_table_p1}.

As indicated in Sect. \ref{sec_HR_spectro}, we could monitor 51 targets with ARCES on the 3.5-m ARC telescope\footnote{Actually, we monitored 53 targets with ARCES but one (KIC 11551404) is not part of the sample of 4465 RGs because its GAIA radius is too small ($2.7\pm0.2 R_\odot$). We picked it before \citet{Berger_2018} was published because it displayed strong rotational modulation and was part of earlier Kepler RG catalogs. The RV dispersion clearly indicates it is a close binary system. We also observed KIC 3459199 which actually has an EB signal in its light curve. It is also an SB. We decided to keep these two in Table \ref{table_ze_table_p1} for the records.}, and make use of 34 archived RV values from the APOGEE/SDSS IV survey\footnote{347 RGs displaying rotational modulation were actually available from the APOGEE database, but only 36 were observed more than three times over a time span longer than 10 days, and 4 were already part of our ARCES sample. Two more were observed only twice -- both are non-oscillating -- but it was enough to prove they are SBs. }. 
For ARCES, we selected the brightest targets among the active ones to maximize the number of observable stars, and we made sure that a significant fraction of the targets did not show any oscillations. Overall, our spectroscopic sample is composed of 85 RGs with clear rotational modulation, including 30 with no oscillations and 55 with oscillations. The result appears to be very different depending whether oscillations are detectable (Fig. \ref{fig_frac_SB_active}). Among the 30 non-oscillating stars, 26 are clear SBs, i.e. $\sim85\,\%$. In contrast, among the 55 oscillating ones, only 8 or 9 are SBs, i.e., $\sim15\,\%$.

\begin{figure}[t!]
\includegraphics[width=8.5cm]{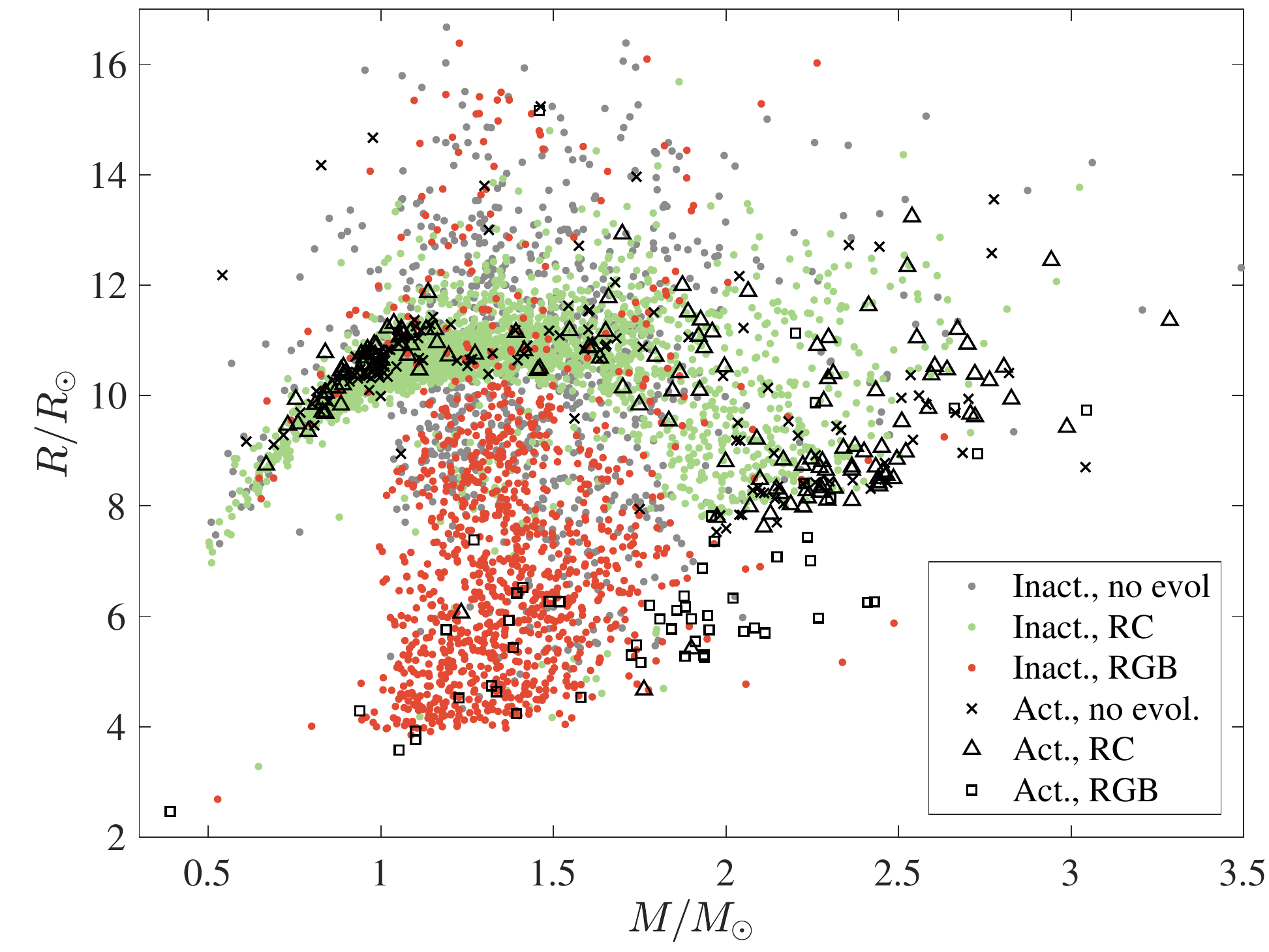} \\
\includegraphics[width=8.5cm]{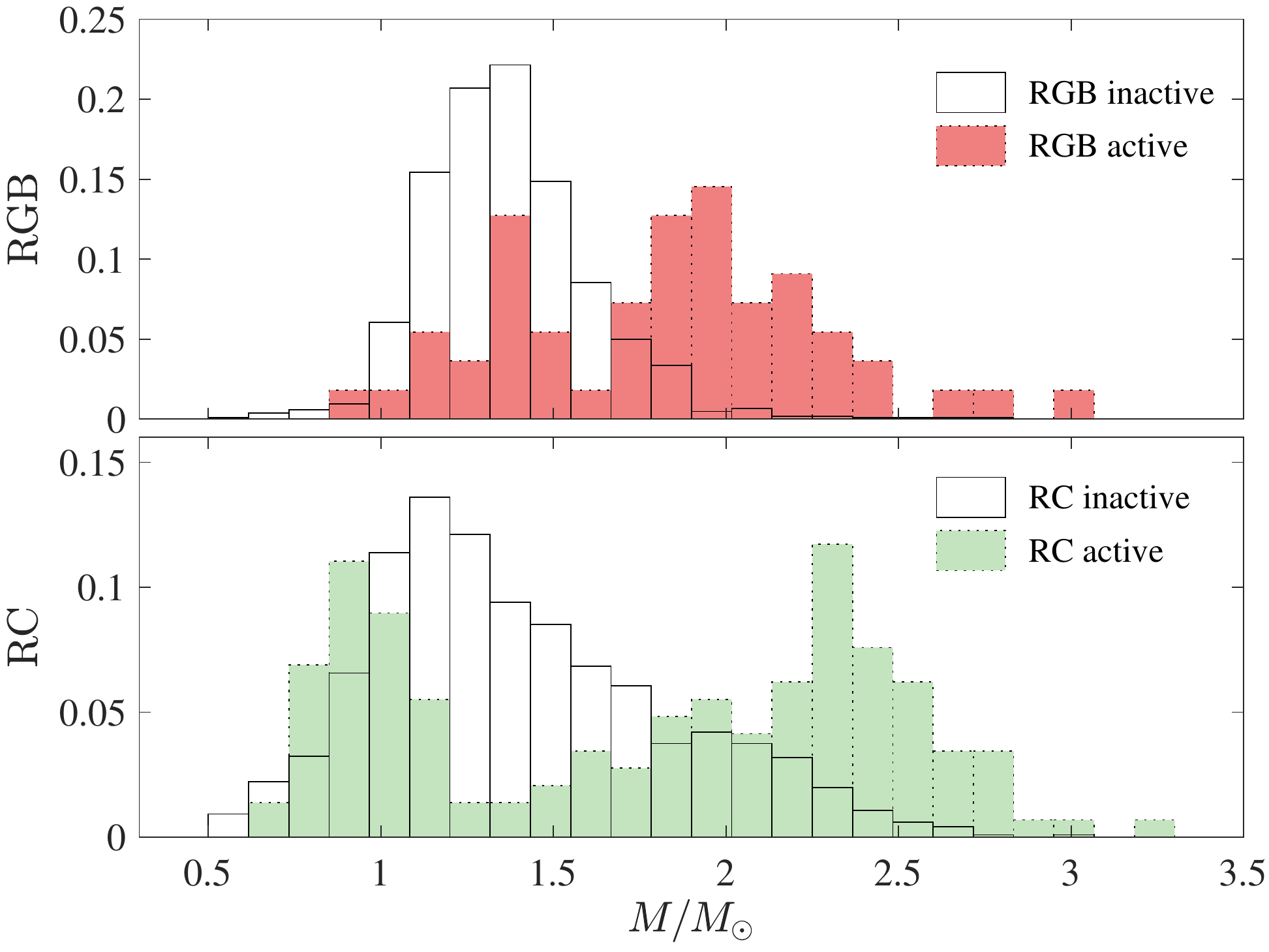} \\
\includegraphics[width=8.5cm]{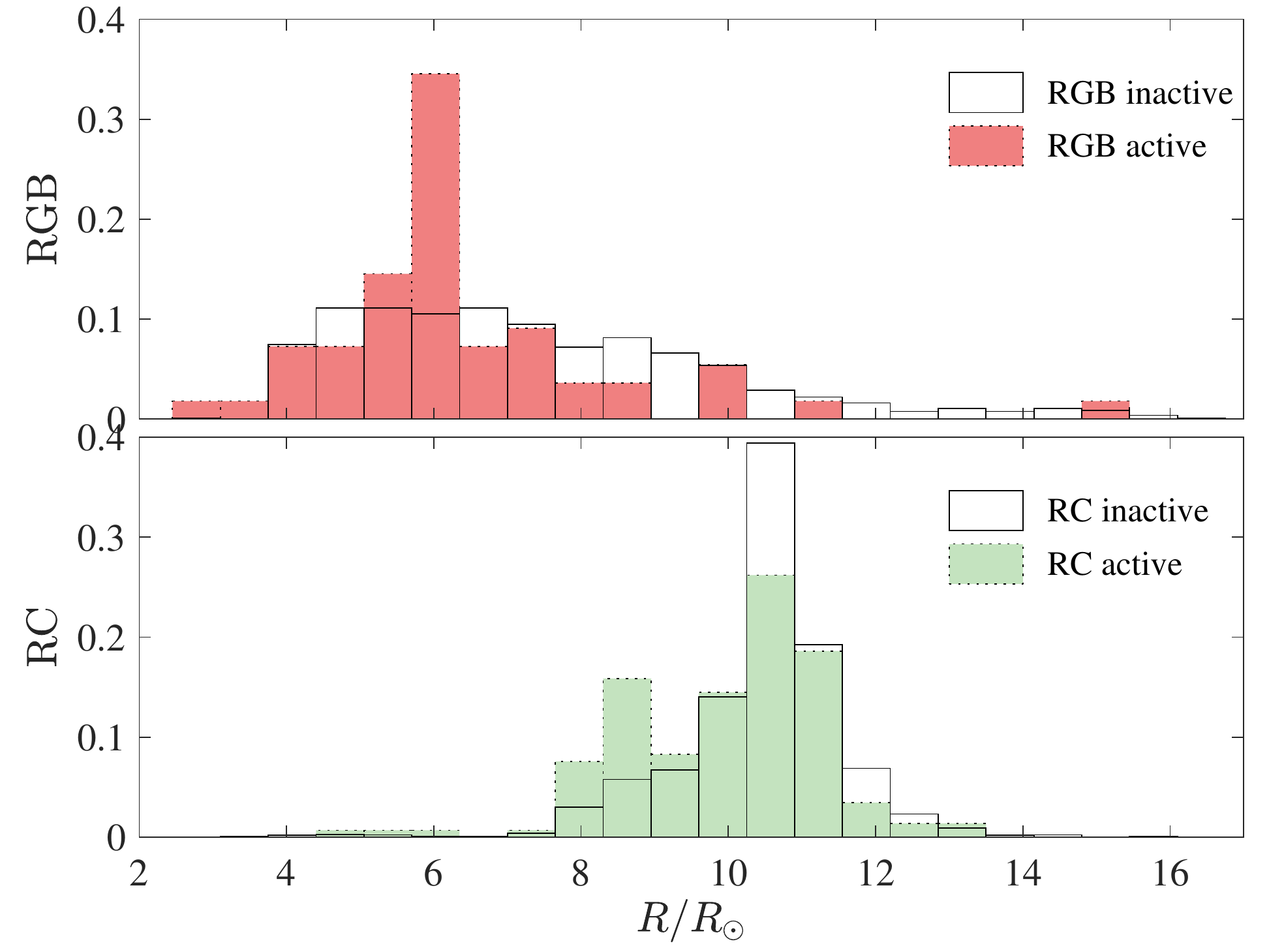} 
\caption{Top panel: radius versus mass of the sample of 4465 stars. Red dots indicate the inactive RGBs, green the inactive RCs, and gray the inactive with unknown evolutionary status. Black squares indicate active RGBs, triangles active RCs and $\times$ signs active RGs with unknown evolutionary status. Panels (b) and (c) represent the distributions of masses and radii of the inactive and active RG samples as a function of stellar masses (top two panels) and radii (bottom two), according to their evolutionary stages. Empty histograms correspond with the inactive samples, red with the active RGBs, and green with active RCs.}
\label{fig_hist_M_R}
\end{figure}
\begin{figure}[t!]
\includegraphics[width=9cm]{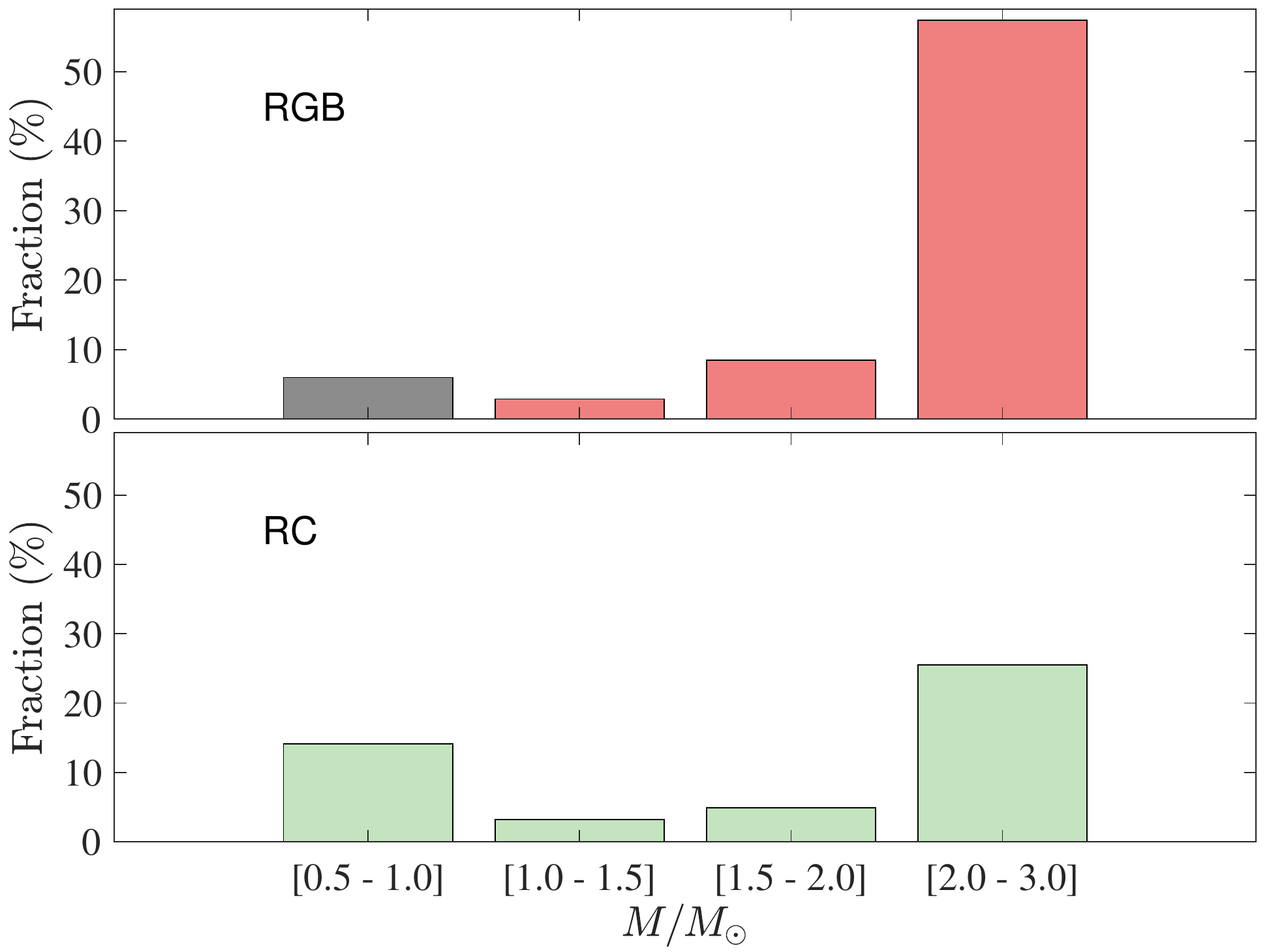} 
\caption{Bar graph showing the fraction of active RGs among the RGB and RC groups as a function of mass. For each mass bin, the estimate of the fraction of active stars is extracted from the number of active stars, the number of inactive stars, and the ratio of stars with a known evolutionary status as explained in Sect. \ref{sec_frac_act_mass} (Eq. \ref{eq_frac_act}). All the numbers used to produce this graph are displayed in Table \ref{table_numbers}. For RGBs with $M \in [0.5, 1.0]\,R_\odot$, the bar is gray because it relies on a very small sample (1 active and 27 inactive RGBs, see Table \ref{table_numbers}), and is not statistically secure.  }
\label{fig_frac_active}
\end{figure}

To check the significance of the SB fraction, we looked for RV data of a null sample, where no rotational modulation is observed. A total of 212 RGs observed more than four times are available in the APOGEE DR16 archive. Among them two are SBs, that is about 1\,\%. We therefore consider the enhanced fraction of close binaries among the RGs with rotational modulation (and weak oscillations) to be significant.

We are aware that the incidence of binarity is often reported to be over 50\,\% \citep[e.g.,][]{Eggleton_2006}. However, this number includes a broad variety of systems, from contact binaries with orbits as short as 0.2 days to wide ones with periods of thousands of years. Since we focus on binaries with at least one member on the RG phase with orbital periods between about 15 to 200 days, our null sample is not representative of the absolute incidence of binaries. 
The catalog of approximately 3,000 EBs discovered by Kepler \citep[e.g.,][]{Prsa_2011,Kirk_2016} helps understanding the order of magnitude of the rate of binaries that we find. The orbital periods of Kepler EBs range from about 0.1 to 1100 days. Only 20\,\% of them have orbits longer than 15 days and about 50 systems have an RG component \citep{Gaulme_Guzik_2019}, that is $\sim2\,\%$ of the total. By assuming a 50\,\% rate of binaries we thus expect about 1\,\% of SBs with an RG component and an orbit longer than two weeks.



\begin{figure*}[t!]
\center
\includegraphics[width=6cm]{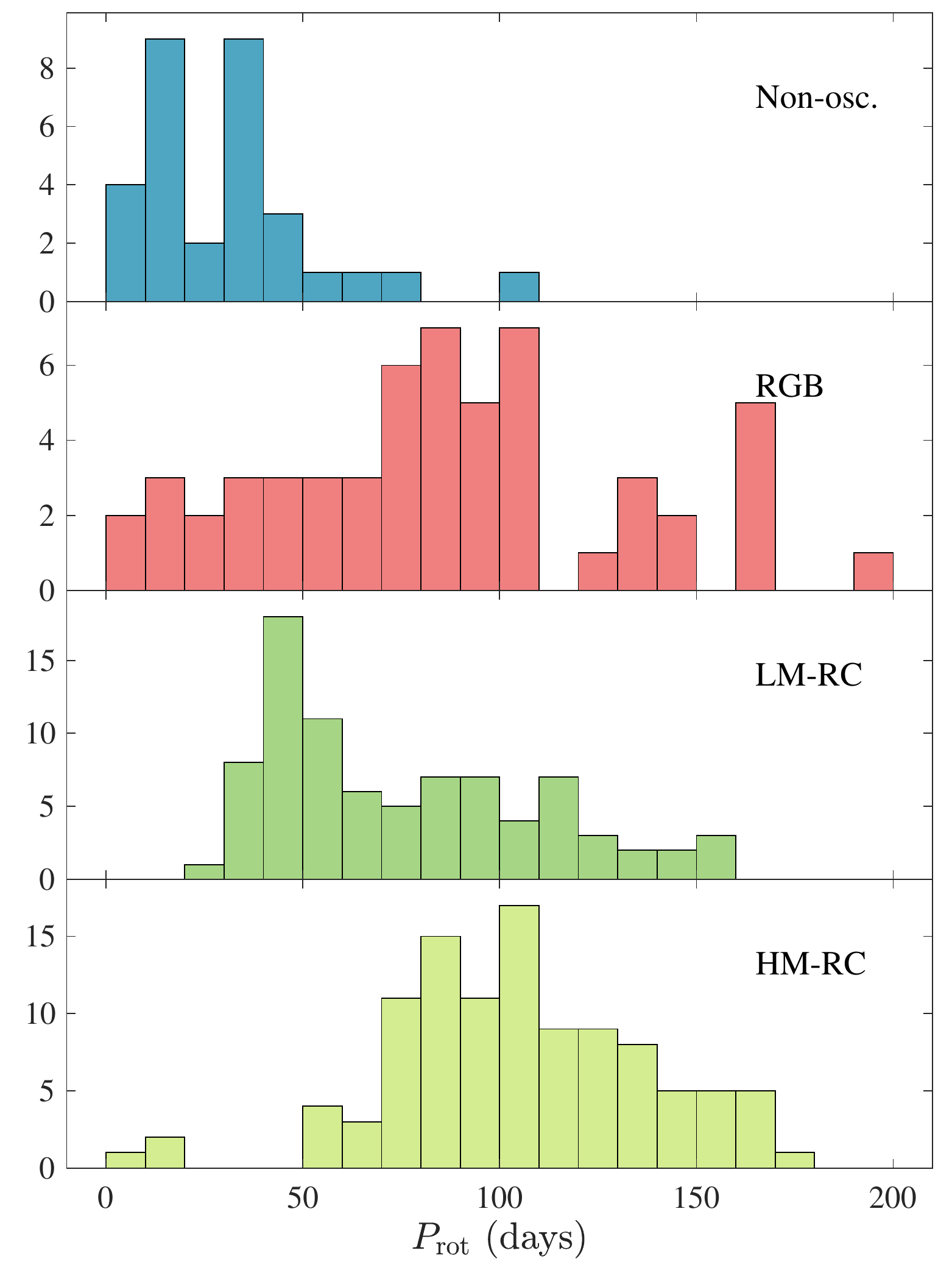} 
\includegraphics[width=6cm]{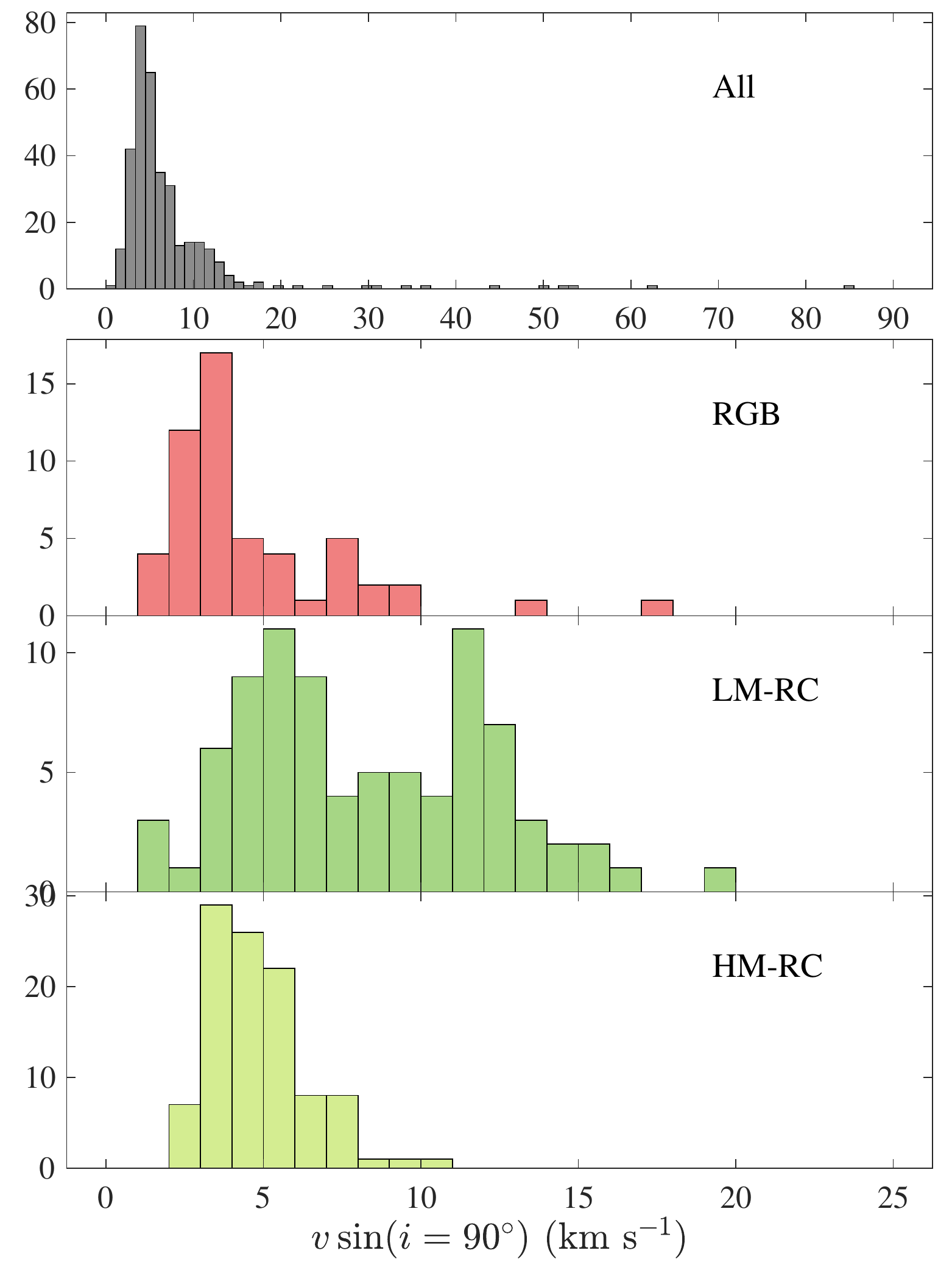} 
\includegraphics[width=6cm]{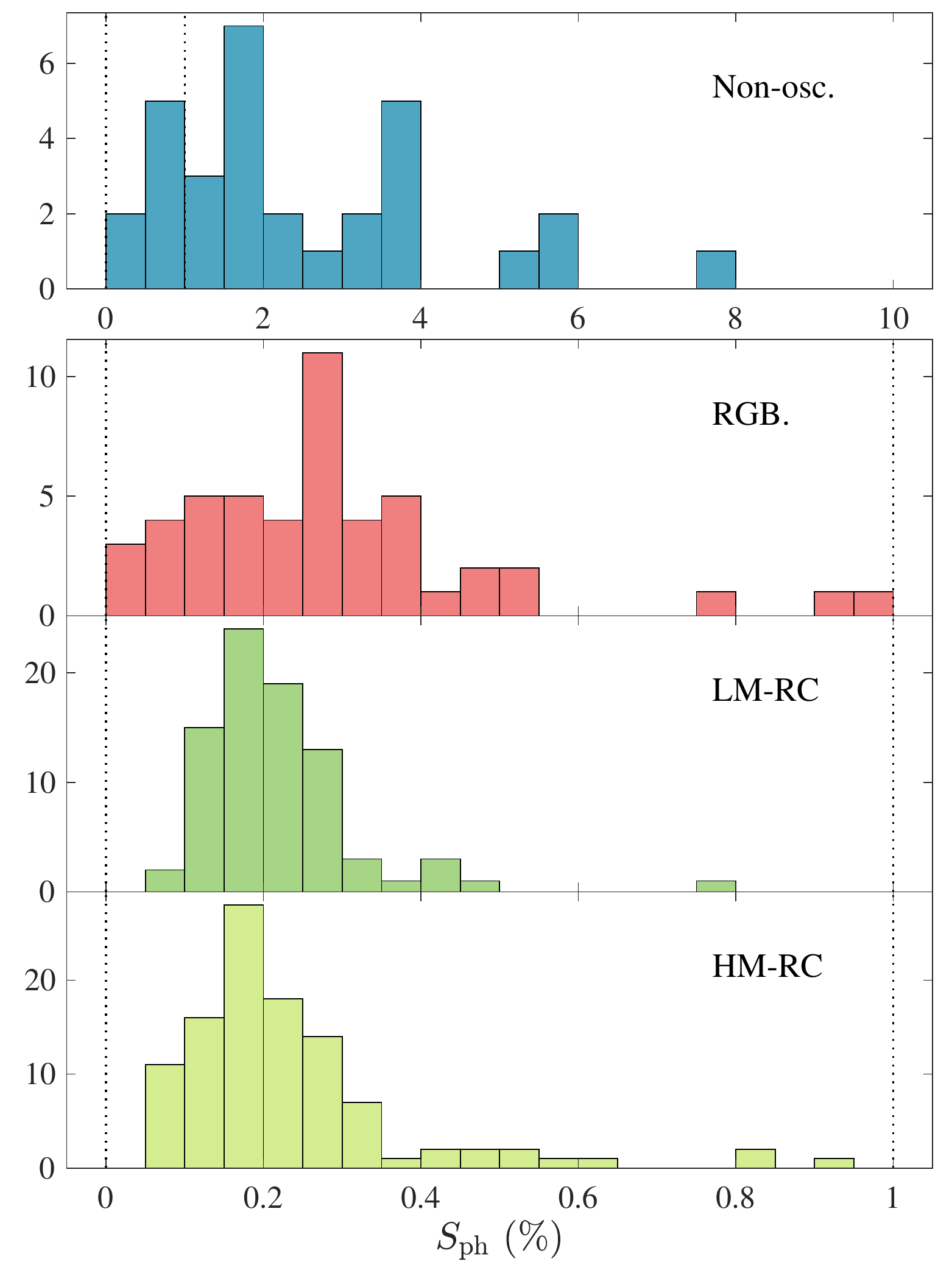} 
\caption{Left panel: distribution of the measured rotation periods expressed in days. Colors indicates RC (green), RGB (red), and non-oscillating (blue) targets. Middle panel: equatorial velocities of the active RGs, that is $v \sin(i=90^\circ$). Top sub-panel: $v \sin(i=90^\circ$) for the whole active sample where oscillations are detected. Next sub-panels: same for RGB, LM-RC, and HM-RC, by considering only the range $[0,25]$ km s$^{-1}$ with bins of 1 km s$^{-1}$. Right panel: distribution of the light-curve standard deviations of the same groups as on left panel, expressed in percent. The dotted lines indicate the range $[0.1,1]\,\%$.}
\label{fig_hist_Prot_etal}
\end{figure*}
\subsection{Rapid rotation without close binarity}
\label{sect_disc}
Among the RGs with rotational modulation, it thus appears that about 85\,\% of the active RGs with detectable oscillations are not part of close binaries. With the help of additional information, we shed light on the origin of their activity.

\subsubsection{Two groups of single active RGs}

We retrieve the masses, radii and evolutionary status of the giants thanks to an asteroseismic analysis described in Sect. \ref{sec_ppties_sample}. By evolutionary status, we mean the RGB where stars burn H in shell around an inert He core, versus the RC, where He burning occurs in the core. 
We distinguish the red clump 1 (RC1) which consists of lower-mass stars ($M\lesssim 2 M_\odot$) that reached the clump after the He flash, from the red  clump 2 (RC2) composed of intermediate-mass stars that smoothly kicked off the He combustion \citep[e.g.,][]{Mosser_2014}. We were able to measure the masses and radii of the 99.3\,\% of RGs with detected oscillations, the evolutionary status of 3205 out of the 4077 inactive RGs, and 200 out of the 340 active RGs (including clear and ambiguous detection of rotational modulation). The inactive sample with a known status is composed of 33\,\% of RGBs and 67\,\% of RCs. The active sample is split in a similar way, with 28\,\% of RGBs and 72\,\% of RCs (see Table \ref{table_numbers}). 

Figure \ref{fig_hist_M_R} shows the distributions of mass and radius for both the inactive and the active samples. A striking result is the very different mass distribution between the two samples (radius distributions are more similar). For the inactive RGs, the mass distribution is consistent with previous reports based on CoRoT or Kepler observations \citep[e.g.,][]{Miglio_2013}. We note that it peaks at $1.3 M_\odot$ for RGBs and $1.15 M_\odot$ for RC, which could be compatible with a significant mass loss during the late stages of the RGB phase. It also could be explained by the fact that we employed a unique asteroseismic scaling relation, whereas they should not be exactly calibrated in the same way for RGB and RC stars \citep{Miglio_2012,Khan_2019}. Regarding the active sample, we first note a general depletion of stars from 1.2 to 2 $M_\odot$. In particular, the RGB sample appears to be depleted of low-mass stars and overpopulated of intermediate mass stars. The RC population is bimodal with a large group of RC1 stars on a narrow range (0.8-1.1 $M_\odot$), and a large group of (mostly RC2s) on a broad range (1.8-2.8 $M_\odot$). Hereafter we often distinguish two groups by referring to low-mass RGBs and RCs (LM-RCs and LM-RGBs) with $M\leq 1.5 M_\odot$, and high-mass RGBs and RCs (HM-RGBs and HM-RCs) above.

\subsubsection{Fraction of active RGs as a function of mass}
\label{sec_frac_act_mass}

If we consider the RGs with unknown evolutionary status to be equally distributed among the RGBs and RCs, we can estimate the absolute fraction of RGs in a given evolutionary status that display surface activity. We note them $r\ind{RGB,act}$ and $r\ind{RC,act}$. For instance, for the RGB group:
\begin{equation}
    r\ind{RGB,act} = \frac{\D{\frac{N\ind{RGB,act}}{r\ind{act,status}}}}{\D{\frac{N\ind{RGB,act}}{r\ind{act,status}} + \frac{N\ind{RGB,inac}}{r\ind{inac,status}} }},
\label{eq_frac_act}
\end{equation}
where the fraction of active RGs with a known evolutionary status $r\ind{act,status} = N\ind{act,status}/N\ind{act} = 200/340 = 58.8\,\%$, and the fraction of inactive RGs with a known status is $r\ind{inac,status} = 3205/4077 = 78.6\,\%$ (see Table \ref{table_numbers}). That way, 6.6\,\% of RGBs and 8.2\,\% of RCs are active.
However, we observe a strong variation of the occurrence of active RGs as both a function of mass and evolutionary status (Fig. \ref{fig_frac_active} and Table \ref{table_numbers}). 

Let us first focus on the low mass part of the sample ($M\leq1.1 M_\odot$) that was considered by \citet{Tayar_2015} and \citet{Ceillier_2017}. Consistently with these two papers, we estimate the fraction of rapidly rotating RGs to be $r\ind{act} = 10.5\,\%$. Our study brings new information thanks to the evolutionary status: it arises that only 3.4\,\% of the RGBs are active, against 11.7\,\% of the RCs. The fact that the fraction of low-mass active RGBs is two to three times smaller than that of active RCs is a strong indicator that the observed rotational modulation originates from an increased rotational rate happening on the RGB. Engulfment or merging of a stellar or substellar companion is a likely hypothesis \citep{Simon_Drake_1989}.


As regards stars with $M\geq2 M_\odot$, we find that about $60\,\%$ of the RGBs and $25\,\%$ of the RC2s are active. First of all, the smaller fraction of active RCs with respect to RGBs is compatible with loss of angular momentum starting during the RGB. Secondly, such a large rate of active RGs among intermediate-mass RGs was not reported by \citet{Ceillier_2017}. As commented earlier, they likely missed this fraction of the sample because of catalog selection biases. 
At the same time, we confirm the conclusions of \citet{Tayar_Pinsonneault_2018} who report that there is no large fraction of rapid rotators among intermediate mass RGs according to spectroscopic $v\sin i$ measurements if we stick to the 10 km s$^{-1}$ criterion. 
Indeed, by combining the rotation periods with the asteroseismic radii, we compute the rotation velocities at equator $v \sin(i=90^\circ)$. Figure \ref{fig_hist_Prot_etal} (middle panel) shows $v \sin(i=90^\circ)$ for the whole sample and for the individual evolutionary stages. Regarding the HM-RC stars, only $\approx 4\,\%$ show $v \sin(i=90^\circ) \geq 10$ km s$^{-1}$.

Interestingly, we notice that the only group that shows a significant fraction ($\approx 37\,\%$) of stars with  $v \sin(i=90^\circ) \geq 10$ km s$^{-1}$ is the LM-RC group, which is suspected to have engulfed a stellar or substellar companion. Finally, by considering all of our RGs with detected rotational modulation and oscillations, the fraction of stars with $v \sin(i=90^\circ) \geq 10$ km s$^{-1}$ is about 17\,\%. By assuming that all of the inactive RGs have $v \sin i < 10$ km s$^{-1}$, it means that 1.4\,\% of our whole sample show $v \sin(i=90^\circ) \geq 10$ km s$^{-1}$, which is a little less than the 2\,\% reported by \citet{Carlberg_2011}.

\begin{figure}[t!]
\includegraphics[width=8.5cm]{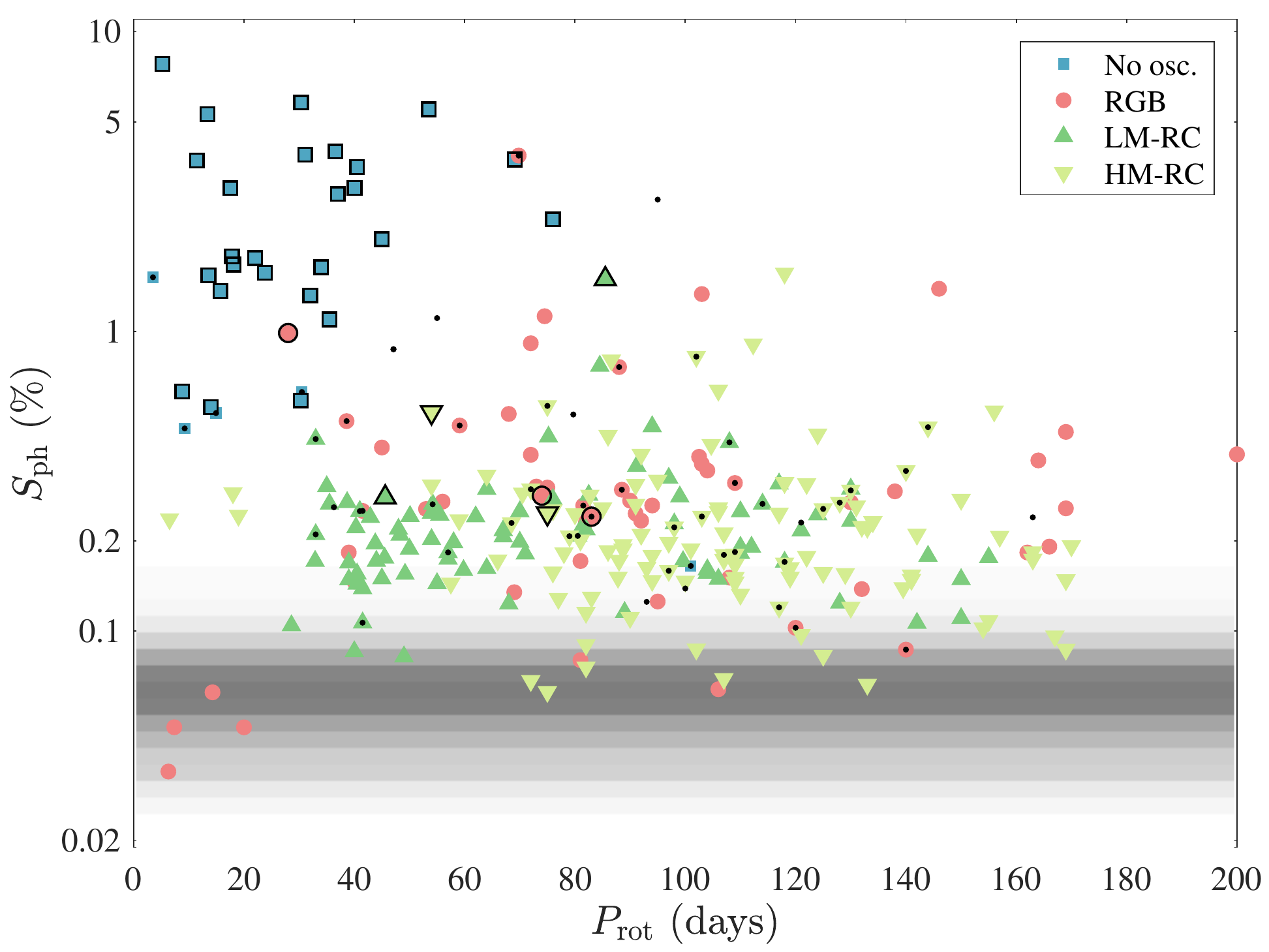} 
\caption{Photometric index $S\ind{ph}$ (percent) as a function of rotation period $P\ind{rot}$ (days) for the RGs with rotational modulation. The photometric index $S\ind{ph}$ is plotted in log scale, $P\ind{rot}$ in linear scale. Blue squares indicate the non-oscillating RGs, red disks the RGBs, darker green upward-pointing triangles the low-mass RCs, and lighter downward-pointing triangles high-mass RCs. The markers that display black edges are confirmed spectroscopic binaries. Black dots indicate RV stable RGs. The gray background reflects the distribution of $S\ind{ph}$ of the inactive stars (the darker, the higher). }
\label{fig_Sph_Prot}
\end{figure}
\begin{figure}
    \centering
    \includegraphics[width=0.495\textwidth]{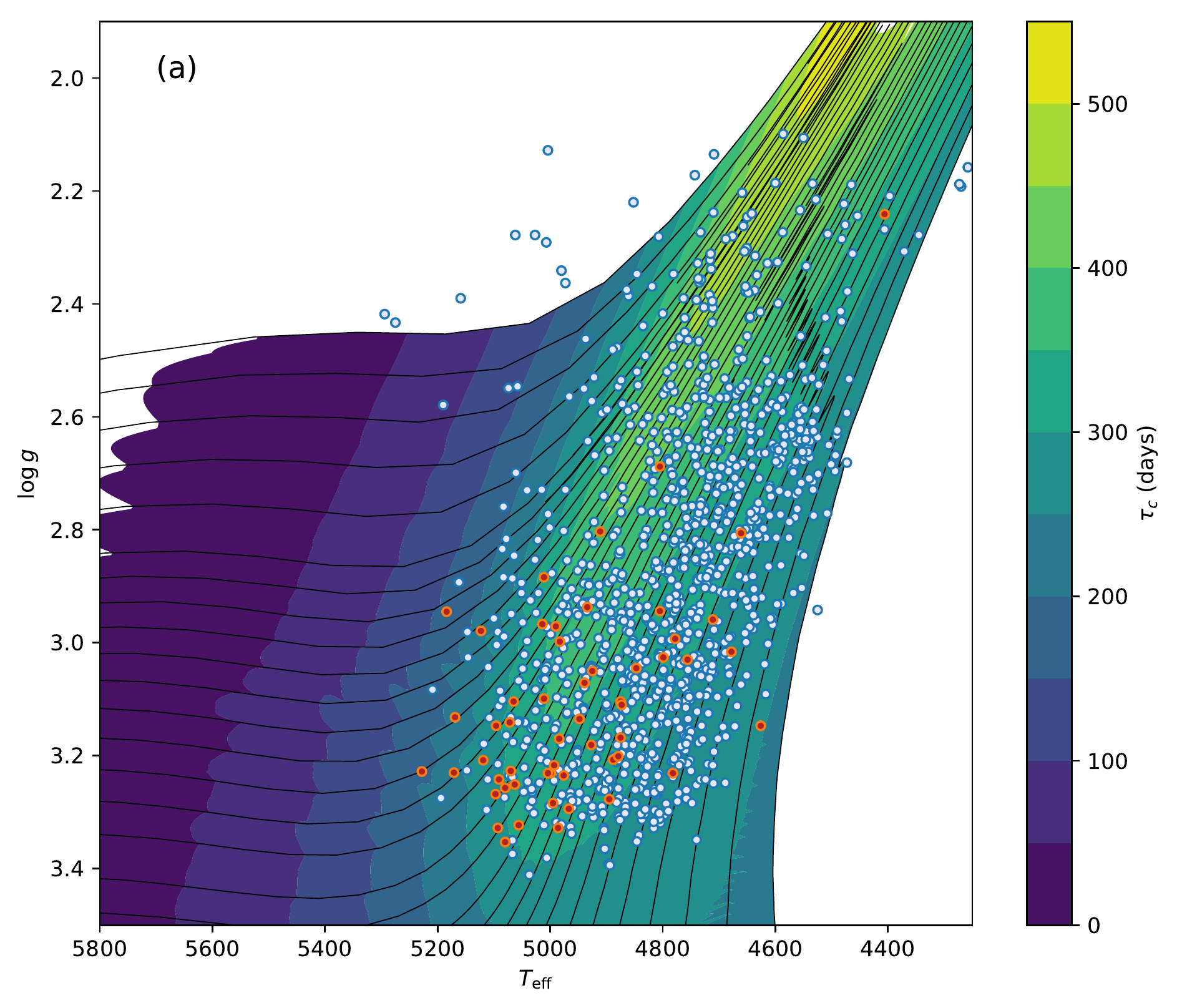}
    \includegraphics[width=0.495\textwidth]{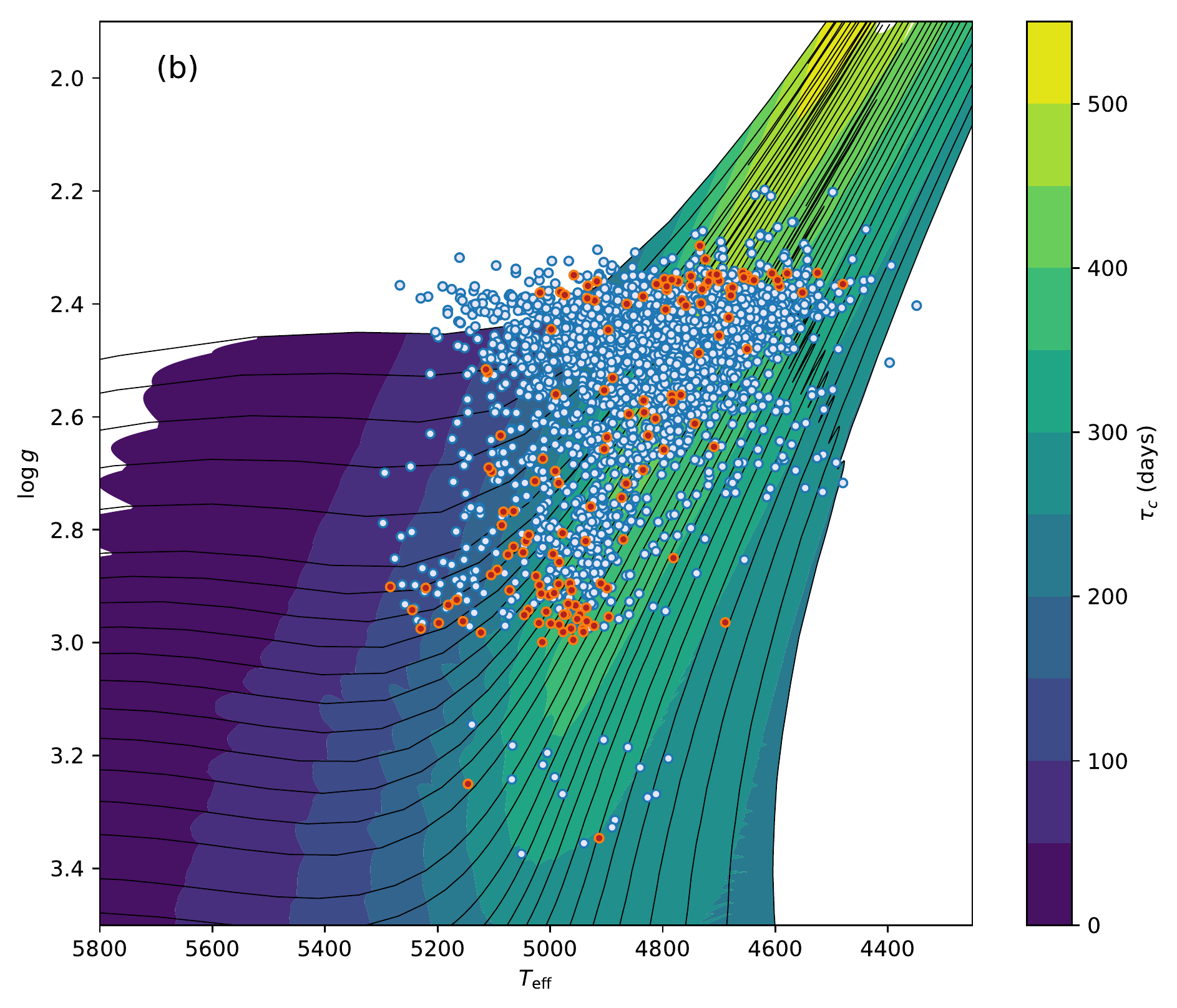}
    \caption{Stars with detected oscillations in our sample compared with stellar evolution tracks and isocontours of the convective turnover timescale in the $(T_{\rm eff},\log\,g)$ plane.
    Panel (a): RGB stars; panel (b): RC stars; stars with and without detected rotational modulation (i.e., ``magnetic activity'') are plotted in red and blue, respectively.}
    \label{fig_tauc_hrd}
\end{figure}
\subsubsection{Rotation rate versus activity}

Firstly, we observe a strong difference of the photometric activity index $S\ind{ph}$ between the non-oscillating (almost all close binaries) and the oscillating active RGs (mostly single RGs) from our sample. From Figs. \ref{fig_hist_Prot_etal} and \ref{fig_Sph_Prot}, the non-oscillating RGs show values of $S\ind{ph}$ from about 1 to 10\,\%, whereas all of the others display maximum values an order of magnitude lower (0.1 to 1\,\%). 
As shown by \citet{Gaulme_2014}, \citet{Gaulme_2016a}, and Benbakoura et al. (submitted), the non-oscillating RGs in EBs are almost all both synchronized and circularized. The few non-locked systems are pseudo-synchronized. It means that the rotation rate is not the only factor driving the amplitude of rotational modulation and oscillations: tidal locking causes in a way or another more photometric contrast. 
Understanding that aspect goes beyond the scope of this paper.

Regarding the single active RGs, our results show that rotational modulation appears at rotation rates much smaller than the ``fast rotation'' threshold that is usually considered. We indeed detect rotational modulation for RGs with $v\sin i \geq 2$ km s$^{-1}$. 
To put this result in its usual theoretical context \citep[e.g.,][]{Noyes_1984,Wright_2011}, we consider the generation of magnetic field in the frame of the turbulent dynamo scenario \citep[e.g.,][]{Charbonneau_2014}. In such a scenario, the efficiency of the magnetic field generation is characterized by the Rossby number $\mathrm{Ro}= P\ind{rot}/\tau\ind{c}$. The efficient dynamo regime, and hence surface magnetic fields and spots, is associated with $\mathrm{Ro}\leq 1$.
We thus need an estimate of the convective turnover timescale $\tau\ind{c}$ for the stars in our sample.

In Fig. \ref{fig_tauc_hrd}, isolevels of $\tau\ind{c}$ are overplotted with the data; evolutionary tracks are also shown for reference. We use the $\tau\ind{c}$ extracted from the stellar evolution tracks from the YaPSI database \citep{Spada_2017}. It should be noted that the values of $\tau\ind{c}$ shown in the Figure were calculated using the mixing-length treatment of convection (see \citealt{Spada_2017} for details), which is adequate to provide an estimate to the leading order.

From Fig. \ref{fig_tauc_hrd}, it is apparent that all of our $\approx 4500$ RGs could be active if their rotational periods were less than about 250 days. In this $T\ind{eff}$-$\log g$ diagram (Kiel diagram), active and inactive RGs overlap, which tends to indicate that the inactive stars rotate slower, likely with rotation periods longer than about 250 days.
Among the RGB stars (panel a) the active sub-sample is concentrated around the region of the first dredge-up, i.e., near the base of the RGB.
For the RC stars (panel b of Figure \ref{fig_tauc_hrd}), two distinct sub-populations of active stars exist, roughly corresponding to the low-mass and the intermediate mass regimes. 
The former have $\log g\approx 2.4$, while the latter have larger surface gravity and are distributed with approximately the same scatter in $\log g$ and $T\ind{eff}$. Both characteristics are consistent with the predictions of classic stellar evolution theory for the RC and the secondary RC, respectively \citep[see, e.g.,][]{Girardi_2016}.

This result also needs be compared with spectro-polarimetric observations of magnetic RGs \citep[e.g.,][]{Auriere_2011, Auriere_2015, Konstantinova_2012, Konstantinova_2013} and the subsequent theoretical work by \citet{Charbonnel_2017}. \citet{Auriere_2015} describes spectropolarimetric observations of a sample of 48 magnetically active RGs, and report clear detection of magnetic fields in 29 stars. Rotation periods are known for most of these stars, and range from about 6 to 600 days. They observe that the magnetic RGs are distributed along a so-called ``magnetic strip'' in the Hertzsprung-Russell diagram, which is a region where $\tau\ind{c}$ is maximum. Such a condition is met during the first dredge up phase at the bottom of the RGB, as well as for RC stars. This result is fully compatible with ours. The only notable difference with their results is that they identify a correlation between the chromospheric activity and the stellar rotation periods. We do not find such a trend between $S\ind{ph}$ and $P\ind{rot}$ (Fig. \ref{fig_Sph_Prot}). This could be due to the fact that their sample spans a broader range of periods (most of our single RGs show periods from 40 to 170 days). Moreover, the chromospheric index $S\ind{index}$ and the Zeeman measurements are not sensitive to the stellar inclination contrarily to $S\ind{ph}$, which makes trends clearer to appear. 

\citet{Auriere_2015} also identify a few outliers that could be descendent of magnetic Ap stars \citep[e.g.,][]{Auriere_2011, Borisova_2016, Tsvetkova_2019}. These outliers are characterized by an enhanced chromospheric activity $S\ind{index}$ with respect to their peers with similar rotation periods. We identify one star among our single RGs with strong activity that could match this description. KIC 5430224 has $S\ind{ph} = 1.1\,\%$ but is RV stable with no particularly fast rotation ($P\ind{rot} \approx 55$ days). It shows very weak oscillations. 
To the contrary the activity of KIC 8160175 ($S\ind{ph} =2.6\,\%$) seems to be caused by contamination of a nearby star (see Appendix \ref{app_contaminants}) and cannot be counted as a possible Ap descendent.

\section{Conclusions}

The original objective of this work was to determine what fraction of RGs shows activity (rotational modulation), and understand its origin. One of the underlying questions was the role of close binarity in this population, standing upon the fact that RGs in close binary systems ($P\ind{orb} \lesssim 200$ days) have been observed to display strong rotational modulation.

To avoid as much as possible being influenced by observational biases, we carefully selected a subsample of the RGs observed by the Kepler satellite during its original four-year mission. This sample was picked from the \citet{Berger_2018} stellar classification of the Kepler field based on the GAIA DR2. From their sample of RGs, we selected the targets whose light curves should not be limited by the photon noise, meaning that the oscillations of a regular RG should be detectable. This led us to consider the brightest stars ($m\ind{Kep} \leq 12.5$ mag) that were observed the longest (more than 3 years). We added a cut on radii, to make sure that the oscillation range would fall between 15 $\mu$Hz and the sampling cut-off (Nyquist frequency) at $\approx 283\,\mu$Hz. The final sample is composed of 4465 RG stars (Fig. \ref{fig_Venn}). 

Our first result is the clear detection of SL oscillations in 99.3\,\% of the sample, which is a much larger fraction than reported in previous studies \citep[e.g.,][]{Stello_2013, Hon_2019}. We explain this mainly thanks to the fact that our special sample is not affected by photon noise for searching for RG oscillations. We also paid much attention to search for oscillations in low S/N conditions. The second important finding is that between 6.7\,\% (only clear detection) and 8.3\,\% (including low S/N detection) of RGs display rotational modulation, which is much larger than reported by previous studies, in particular by \citet{Ceillier_2017}. We also clearly show that the active RGs present oscillations with lower amplitudes with respect to the inactive sample (Fig. \ref{fig_sample}). This latter fact -- increased activity leads to lower modes -- was previously observed for MS stars \citep[e.g.,][]{Chaplin_2011a,Mathur_2019}, but also for RGs in close multiple systems \citep[e.g.,][and Benbakoura et al., submitted.]{Derekas_2011,Gaulme_2014}.

From high-resolution spectroscopic observations specially conducted with the \'echelle spectrometer of the 3.5-m ARC telescope, as well as public SDSS/APOGEE observations, we could determine the role of close binarity in the origin of the active RGs. It appears that among the 30 non-oscillating active RGs, 26 are clear SBs, one lacks in statistically significant RV variations, and another three are possible SBs. This means that almost all of the non-oscillating RGs belong to close binary systems. Among the rest of the sample, that is, the active RGs with partially suppressed but detectable oscillations, the fraction of close binaries is still significantly larger than in among the inactive ones. We count about 15\,\% of SBs whereas we identify only 1\,\% in a test sample composed of inactive RGs. This also means that a large majority of active RGs with detectable oscillations do not belong to close SBs.

\begin{figure}
    \centering
    \includegraphics[width=0.5\textwidth]{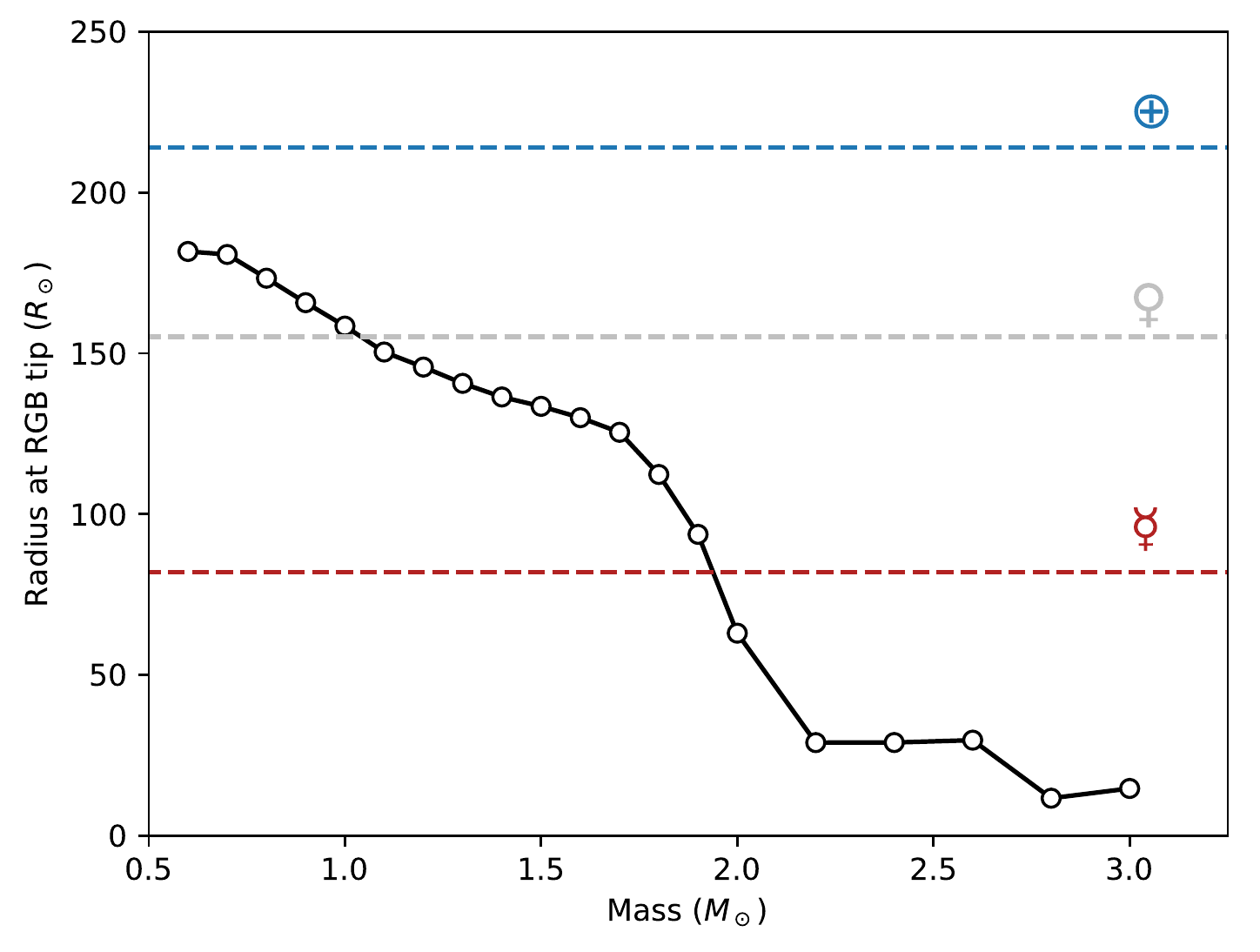}
    \caption{Stellar radius at the tip of the red giant branch as a function of mass as predicted by the stellar models used in this work. No mass loss was considered during the RGB phase. The semimajor axes of the current orbits of Mercury, Venus, and Earth are also marked for comparison.}
    \label{fig_radius_tip_RGB}
\end{figure}

\begin{figure}
    \centering
    \includegraphics[width=0.5\textwidth]{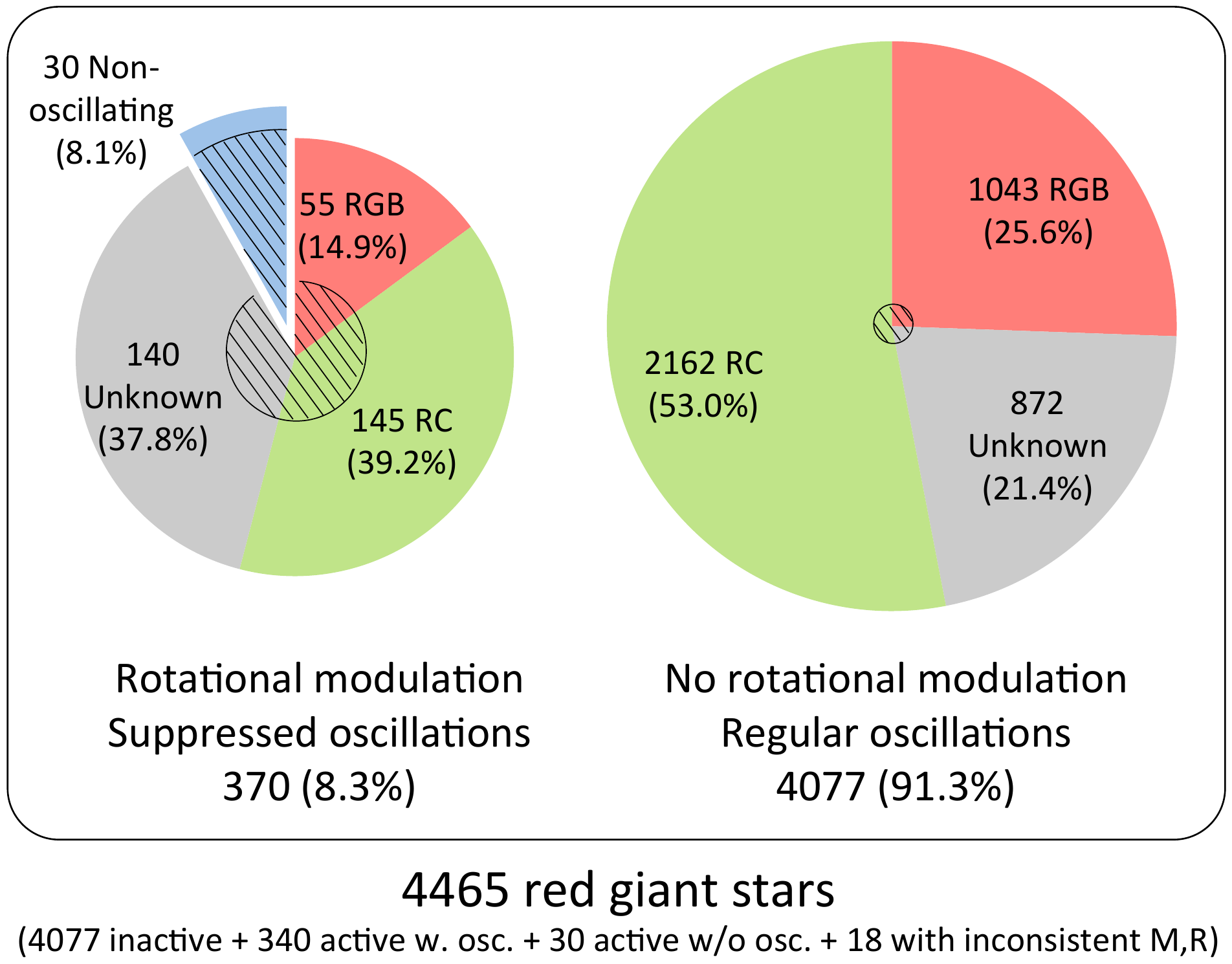}
    \caption{Diagram that summarizes the sample we analyze in this paper. The hatched areas indicate the fraction of spectroscopic binaries among a given population. They represent about 85\,\% of the non-oscillating active RGs, 15\,\% of the active RGs with partially suppressed oscillations, and 2\,\% of the inactive RGs according to our spectroscopic sample. We note that the percentages in the active sample slightly differ from those reported in Table \ref{table_numbers} because they refer to the 370 active stars (with and without detected oscillations) in the present diagram, whereas they refer to the 340 RGs with oscillations in Table \ref{table_numbers}. }
    \label{fig_Venn}
\end{figure}

We shed light on the origin of the active single RGs by analyzing their distributions in terms of physical properties (mass, radius) and evolutionary status (RGB, RC). It arises that most of the active RGs fall into two distict groups: a low-mass ($M\leq1.5 M_\odot$) and an intermediate-mass range ($1.8\leq M\leq2.8 M_\odot$). 
The low-mass part of the active RGs is mostly composed of stars on the RC: about 3\,\% of the RGBs are active while about 12\,\% of the RCs are active. This means that a fraction of the RGs in that mass range gains angular momentum between the RGB and the RC. This observation favors the scenarios of planet engulfment or merging with small stellar objects. Indeed, low-mass stars ignite Helium after the He flash, and present radii up to $\sim200\ R_\odot$ at the tip of the RGB (Fig. \ref{fig_radius_tip_RGB}), and may swallow planets, converting their orbital momentum into spin and friction loss. Another option is a stellar engulfment. Highly eccentric systems ($0.3<e<0.8$) as those composing the heartbeat systems \citep[e.g.,][]{Beck_2014,Kuszlewicz_2019} are typically composed of a Sun-mass star and a much smaller companion (0.2 $M_\odot)$. Near the tip of the RGB, the most massive star has a radius larger than the periastron distance and thus merges with the companion. 

If this scenario of planet or stellar engulfment is true, it is interesting to notice that it regards a relatively narrow mass range. This can be qualitatively explained by three factors. Firstly, since more massive stars ($M\geq2M_\odot$) do not reach the clump through the He flash, their radii do not exceed $\approx20\,R_\odot$, so they are less susceptible to engulf planets. Moreover, because they are more massive, they gain relatively less angular momentum by swallowing planets than less massive stars. Lastly, the orbital angular momentum of close-in planets increases as a function of the semi-major axis, and thus is less efficient with a close planet with respect to a planet like Venus, orbiting at 150 $R_\odot$. 

As regards intermediate-mass stars, we observe that a large fraction of the RGBs is active (over 50\,\% when $M\geq 2M_\odot$), and a large but reduced fraction of active RCs (about 25\,\%). In opposition to the low-mass range, this observation implies that intermediate-mass stars tend to loose angular momentum between the RGB and the clump. Both the fraction of the active RGB in this mass range and the fact that they loose momentum between the RGB and the RC are compatible with the expected scenario for intermediate-mass stars. They do not loose angular momentum during the MS, thus still spin fast on the RGB, and a little less once on the clump. However, the values are still in contrast with recent theoretical studies \citep[e.g.,][]{Tayar_Pinsonneault_2018}. From the observed distribution of rotation rates of A and B-type MS stars, about half of the RGs are expected to display $v\sin i\geq 10 $km s$^{-1}$ \citep[e.g.,][]{Ceillier_2017}. From our asteroseismic radii and photometric rotation periods, we measure that only about $4\,\%$ of the active intermediate-mass RGs display such a rapid rotation, i.e. about 2\,\% of that mass range by including the inactive ones. This reinforces previous reports of an unidentified sink of angular momentum after the MS. 

Regarding the connection between the emergence of surface activity, we observe that RGs with $v\sin i\geq 2 $km s$^{-1}$ show rotational modulation. This is actually not surprising when we compare the rotation periods of the active RGs with the convective turnover time $\tau\ind{c}$. In a turbulent dynamo scenario, the generation of magnetic fields starts being efficient if the rotation period is lower than $\tau\ind{c}$. All the active RGs meet these conditions (Fig. \ref{fig_tauc_hrd}). 
A correlated interesting fact is that the active RGs that belong to binary systems and that do not display oscillations show a photometric activity index $S\ind{ph}$ about an order of magnitude larger than single active RGs. We could explain this by the fact that the non-oscillating active RGs have rotation periods shorter in average than the single active RGs. 
However, we observe that single RGs with rotation periods between 30 and 80 days still show an $S\ind{ph}$ about ten times less than the RGs in SBs with the same periods. This implies that tidal locking somehow leads to larger magnetic fields, and also a more efficient oscillation suppression. Investigating that aspect will be part of future works. 
Last, \citet{Mathur_2019} report that MS stars with $S\ind{ph} \geq 0.2\,\%$ never show SL oscillations. We note that this threshold is not the same for RGs, as we detect oscillations up to levels of about $S\ind{ph} \approx 1\,\%$.

Further developments of this project will cover several aspects. Firstly, complementary high-resolution measurements of the identified SBs will help characterizing better the types of systems that compose the population of close binaries with an RG. In particular, we can infer the mass ratio for the SB2s. We will also study the possibility of leading infrared interferometric measurements to resolve the brightest and closest systems and retrieve the mass and radii of the individual components. 
As regards the other topics covered in this work, we will perform a detailed analysis of the HR spectra and look for chemical peculiarities of the LM-RC stars, in order to spot possible signatures of merging history. Among possible tracers, we will especially monitor the Lithium absorption line as suggested by e.g., \citet[][]{Soares-Furtado_2020}, which we have already identified for a handful of targets (see Table \ref{table_ze_table_p1}). As regards the fast rotating intermediate-mass RGs, we will investigate in more details the analysis of mixed modes to estimate the rotation rate of the core, to measure the evolution of angular momentum between the RGB and RC phases.

\section*{Acknowledgments} 
P. Gaulme, F. Spada, and C. Damiani acknowledge funding from the German Aerospace Center (Deutsches Zentrum f\"ur Luft- und Raumfahrt) under PLATO Data Center grant 50OO1501. M. Vrard was supported by FCT - Funda\c{c}\~ao para a Ci\^encia e a Tecnologia  through national funds and by FEDER through COMPETE2020 - Programa Operacional Competitividade e Internacionaliza\c{c}\~ao in the context of the grants: PTDC/FIS-AST/30389/2017 \& POCI-01-0145-FEDER-030389 and UID/FIS/04434/2013 \& POCI-01-0145-FEDER-007672. M.B. has been funded by the PLATO CNES grant. Much of this paper is based on observations obtained with the Apache Point Observatory 3.5-meter telescope, which is owned and operated by the Astrophysical Research Consortium. The authors also used archived data from the data release DR16 of the APOGEE/SDSS IV survey. Funding for the Sloan Digital Sky Survey IV has been provided by the Alfred P. Sloan Foundation, the U.S. Department of Energy Office of Science, and the Participating Institutions. SDSS acknowledges support and resources from the Center for High-Performance Computing at the University of Utah. The SDSS web site is www.sdss.org. SDSS is managed by the Astrophysical Research Consortium for the Participating Institutions of the SDSS Collaboration including the Brazilian Participation Group, the Carnegie Institution for Science, Carnegie Mellon University, the Chilean Participation Group, the French Participation Group, Harvard-Smithsonian Center for Astrophysics, Instituto de Astrofísica de Canarias, the Johns Hopkins University, Kavli Institute for the Physics and Mathematics of the Universe (IPMU) / University of Tokyo, the Korean Participation Group, Lawrence Berkeley National Laboratory, Leibniz Institut für Astrophysik Potsdam (AIP), Max-Planck-Institut für Astronomie (MPIA Heidelberg), Max-Planck-Institut für Astrophysik (MPA Garching), Max-Planck-Institut für Extraterrestrische Physik (MPE), National Astronomical Observatories of China, New Mexico State University, New York University, University of Notre Dame, Observatório Nacional / MCTI, The Ohio State University, Pennsylvania State University, Shanghai Astronomical Observatory, United Kingdom Participation Group, Universidad Nacional Autónoma de México, University of Arizona, University of Colorado Boulder, University of Oxford, University of Portsmouth, University of Utah, University of Virginia, University of Washington, University of Wisconsin, Vanderbilt University, and Yale University.


\bibliographystyle{aa} 
 \bibliography{bibi.bib}





\appendix 


\section{Expected amplitude of RV measurements}
\label{app_RV_theo}

It is important to know what to expect in terms of RV measurements. From the studies of \citet{Rawls_2016}, \citet{Gaulme_2016a} and Benbakoura et al. (submitted), surface activity of RGs in EBs was observed with pseudo-periods ranging from 20 to 180 days, for systems with orbital periods ranging from about 15 to 120 days. The large majority of them are composed of an RG on the RGB, and an MS star (from M to F types) on low-eccentricity orbits ($e\lessapprox 0.1$). The RG masses $M_1$ range from 0.9 to 1.6 $M_\odot$ with typical mass ratios $q=M_2/M_1$ of 0.7 to almost 1. We note the presence of an outlier, KIC 9246715, which is an an eccentric system ($e=0.36$) composed of a pair of $2.1 M_\odot$ RG stars, likely to be on the horizontal branch \citep[][]{Rawls_2016}. Another outlier to be noticed is KIC 8702921 which displays $q=0.16$. For our simulations, we therefore consider a range of systems with orbital periods from 10 days to 200 days, individual masses from 1 to $2.5 M_\odot$, mass ratios from and 0.1 to 1, and eccentricities from 0 to 0.4. To complete our simulations we consider the archetypal system to have a $M_1=1.2 M_\odot$, $M_2=1.0 M_\odot$, $P\ind{orb}=40$ days, and $e=0$. 

\begin{figure}[t!]
\includegraphics[width=8.5cm]{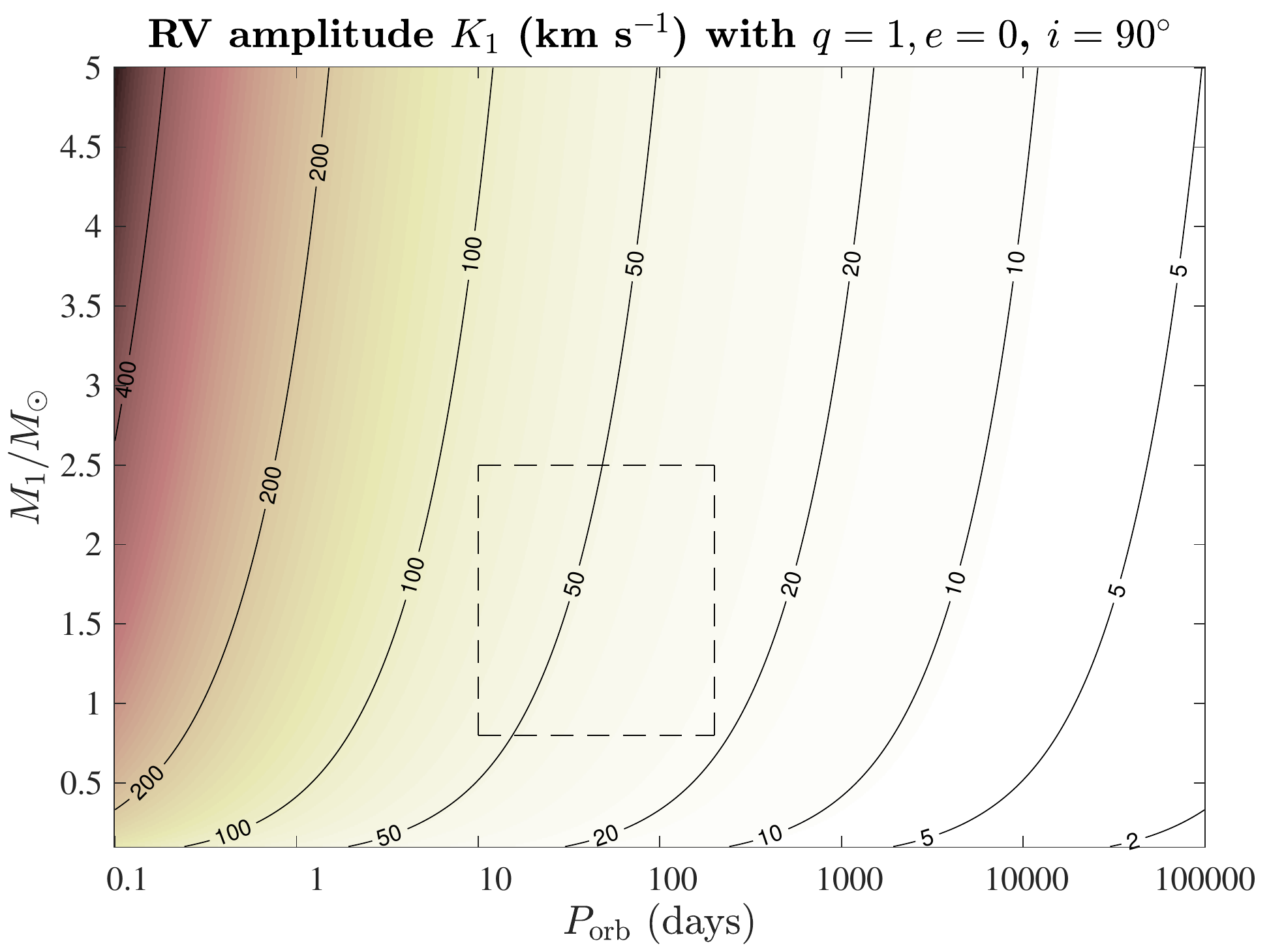} 
\includegraphics[width=8.5cm]{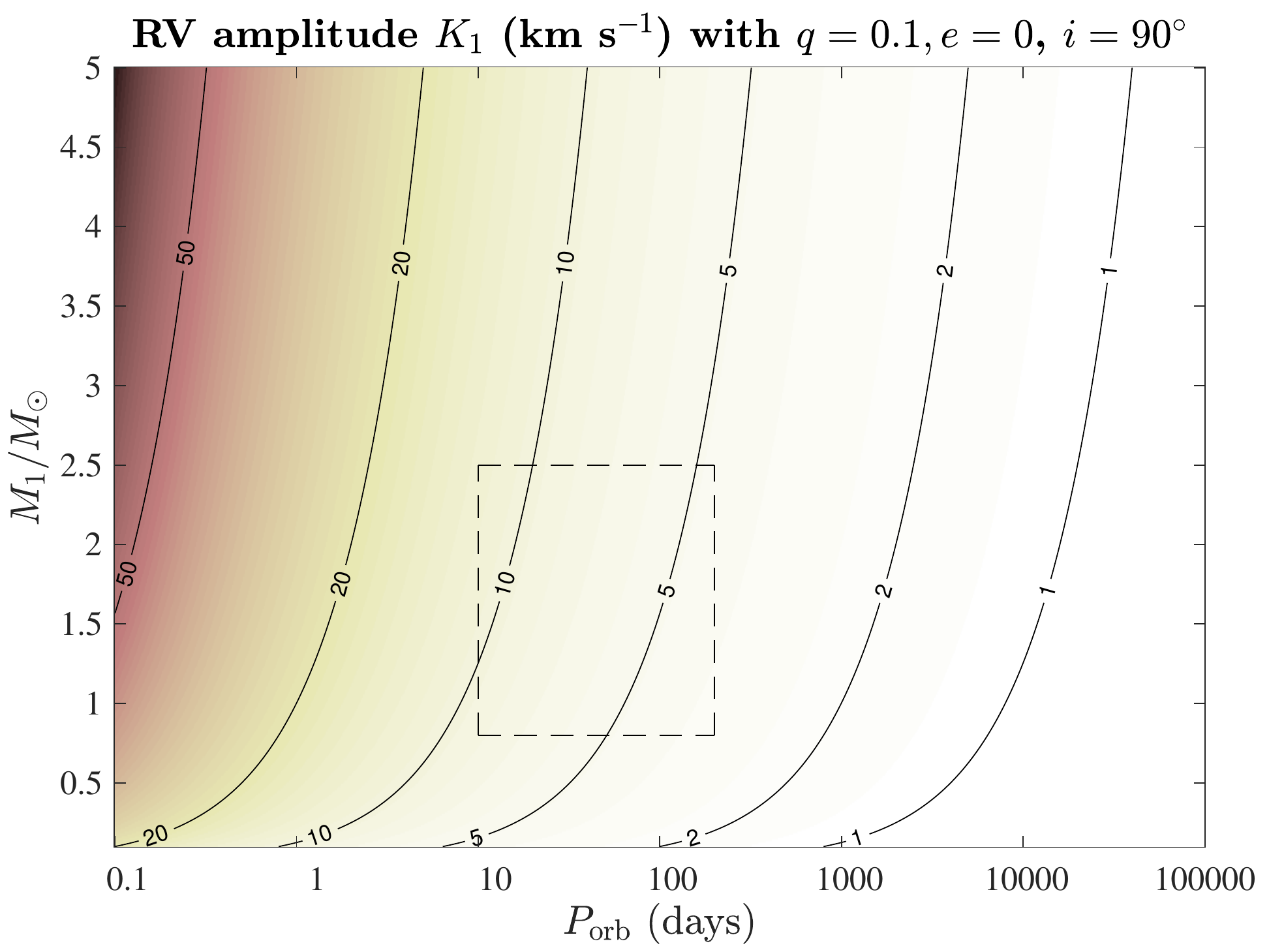} 
\caption{Effect of mass ratio on the RV semi-amplitude $K$ of the RG component. The $x$-axis is the orbital period $P\ind{orb}$ expressed in days, which runs from 0.1 day to $\approx 274$ years. The $y$-axis is the RG mass $M_1$. Systems are assumed to have a zero eccentricity ($e=0$) and to be seen edge-on ($i=90^\circ$). Orbital periods $P\ind{orb}$ are considered from 0.1 to 100,000 days ($\approx 274$ years), and the RG mass $M_1$ from 0.1 to 5 $M_\odot$. Top panel: with a mass ratio $q=1$; bottom: with $q=0.1$. The area delimited by dashed lines indicates the range of values met for systems in which surface activity is measured.}
\label{fig_RV_amplitude_mass}
\end{figure}
\begin{figure}[t!]
\includegraphics[width=8.5cm]{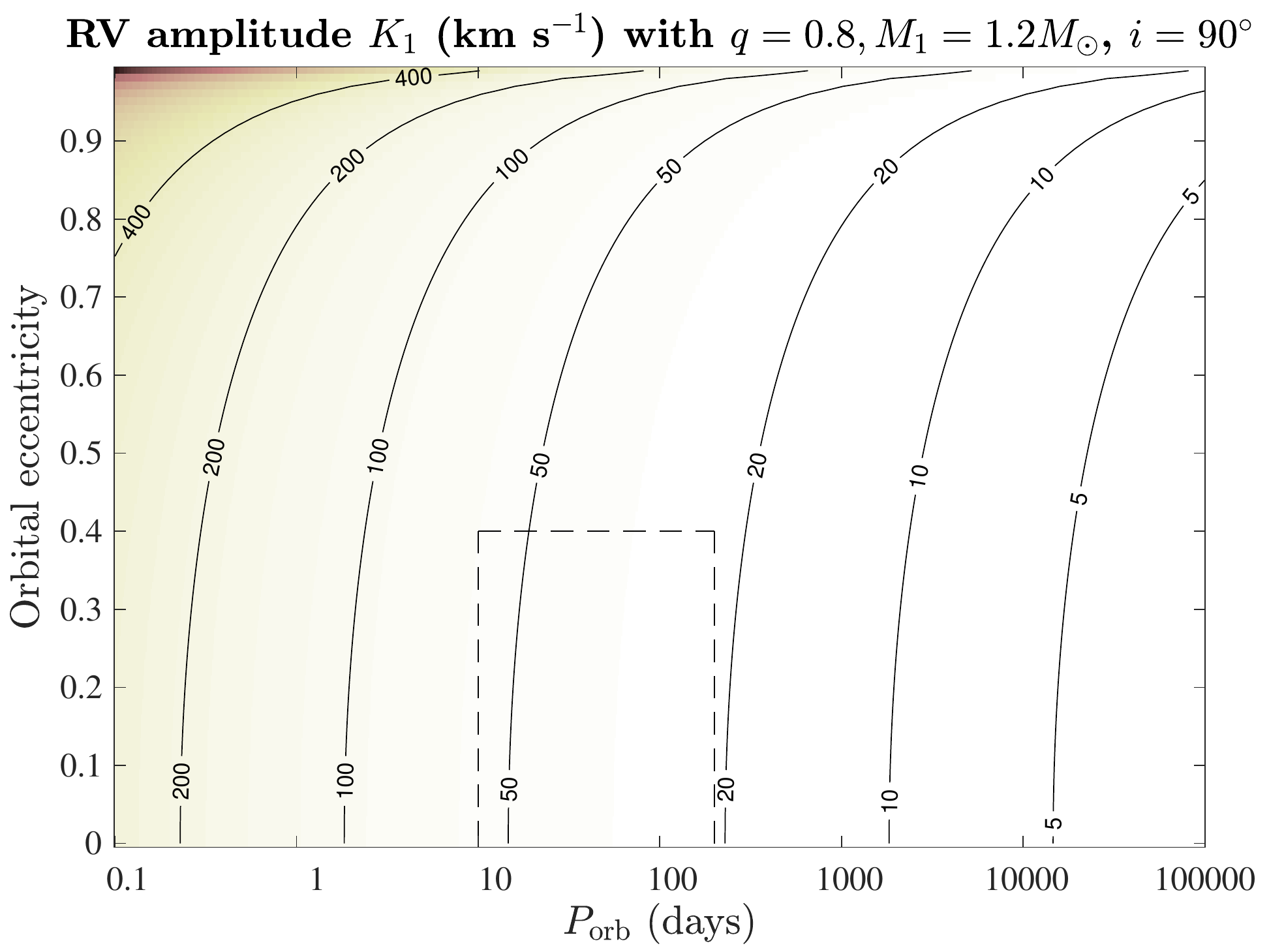} 
\caption{Effect of eccentricity on the RV semi-amplitude $K$ of the RG component. The $x$-axis is the orbital period $P\ind{orb}$ expressed in days, which runs from 0.1 day to $\approx 274$ years. The $y$-axis is the orbital eccentricity. Systems are assumed to have a mass ratio ($q=0.8$) with $M_1=1.2 M_\odot$, and to be seen edge-on ($i=90^\circ$). The area delimited by dashed lines indicates the range of values met for systems in which surface activity is measured.}
\label{fig_RV_amplitude_ecc}
\end{figure}

\begin{figure}[ht!]
\includegraphics[width=8.5cm]{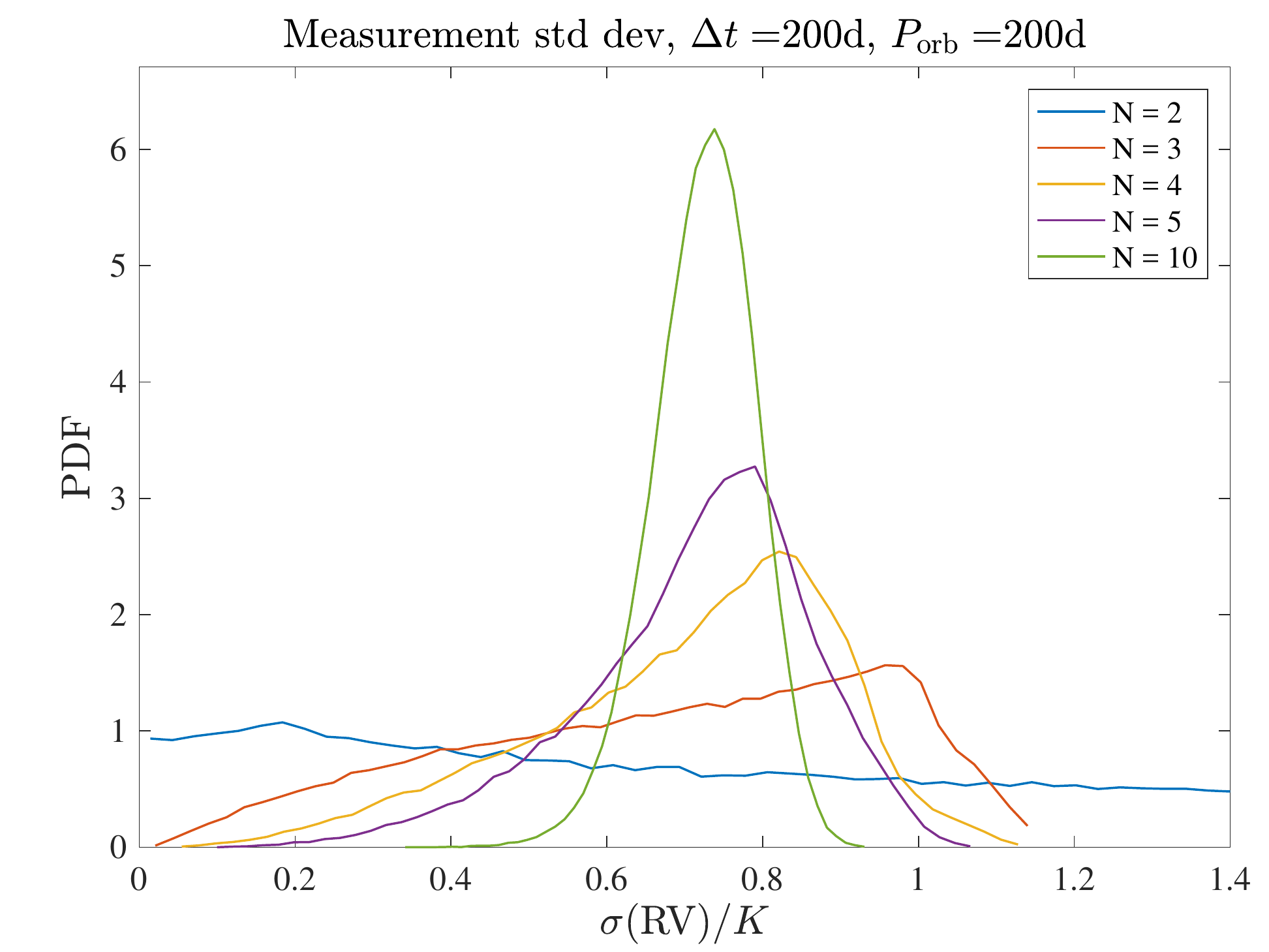} 
\caption{Probability density distribution of the standard deviation of RV measurements for a system with a 200-day orbit. We consider 2, 3, 4, 5, or 10 measurements spaced of at least 7 days realized on a time span of 200 days. Standard deviation is normalized to the RV semi amplitude $K$.}
\label{fig_std_measured_RV}
\end{figure}

We remind that the RV of the RG component is written as:
\begin{equation}
    V_1 = \gamma_1 - K_1 [\cos(\omega + \mathcal{V}) + e \cos \omega];
\end{equation}
where $\gamma_1$ is the barycentric RV, $K_1$ the RV semi-amplitude, $\omega$ the argument of periastron, $\mathcal{V}$ the true anomaly, and $e$ the eccentricity. The semi-amplitude $K_1$ is a function of the stellar masses and orbital properties: 
\begin{equation}
    K_1 = \frac{M_2}{(M_1+M_2)^{2/3}} \left(\frac{2 \pi G}{P (1-e^2)^{3/2}}\right)^{1/3}
\end{equation}
For circular orbits ($e=0$), the RV signal is a sine curve with an amplitude $K_1$. Figure \ref{fig_RV_amplitude_mass} shows the effect of the total mass and mass ratio of a given system on the semi-amplitude $K_1$ of the RV variations of the RG component. In the ranges of systems that we consider, it appears that the typical value of $K_1$ is $\approx30$ km s$^{-1}$, while it can range from about 3 to 90 km s$^{-1}$ in some more extreme cases.
Figure \ref{fig_RV_amplitude_ecc} shows the impact of eccentricity on the values of $K_1$. It appears that it has negligible impact when we consider systems with eccentricities lower than 0.4.

The inclination of the orbital plane $i$ determines the amplitude of the actual RV signal. By assuming a uniform distribution of inclination angles, we note that 90\,\% of the $\sin i$ are larger than 0.16. For example, if a system has a RV semi amplitude peaking at 30 km s$^{-1}$ if observed in its orbital plane, it will display an RV signal peaking more than 4.7 km s$^{-1}$ in 90\,\% of the cases, when observed from a random location. It is thus rare than inclination makes an RV signal totally disappear.

The sampling of the measurements is the last factor to determine the likelihood of detecting an SB. We made a simulation of the RV signal of a system with zero eccentricity and a semi amplitude $K$. We consider the observations happening over a period of 200 days, with at random measurements dates provided that they are separated by at least 7 days from one another. We ran $100,000$ of simulated RV curves with random measurements, by considering either 2, 3, 4, 5 or 10 measurements. As expected, two measurements are statistically not sufficient to detect RV scatter. These simulations led us to observe between 4 and 7 times our targets to assess or not RV variability.


\begin{table}
\tiny
\caption{\tiny Properties of the sample of oscillating RGs. Among the 4465 stars under study in this paper, this table excludes the 30 stars with no oscillations, and another 18 stars with inconsistent $M$ and $R$. Numbers are provided for the whole sample (4417 stars) and for specific mass ranges. We included the $[0.5,1.1]\,M_\odot$ range to compare with results from \citet{Tayar_2015} and \citet{Ceillier_2017}. The ``known status'' column indicates the number and fraction of RGs whose evolutionary status could be determined. The RGB and RC columns report the number and fractions of these types of RGs for each mass range, and the ``$l=1$ depleted'' column indicate the number and fraction of RGs with depleted dipolar modes, according to the \citet{Mosser_2017} criterion. The terms ``active'' and ``inactive'' refer to RGs with or without detected rotational modulation. The label ``frac'' indicates the fraction of active RGBs and RCs among all RGBs and RCs of the considered mass range. These fractions take into account the fraction of known evolutionary status. Most of the content of this table (not the $l=1$ depleted RGs) can be visualized in Fig. \ref{fig_frac_active}.}
\label{table_numbers}
\begin{tabular}{l c c c c c }
\hline 
          & All	       &Known      & RGB        & RC          & $l=1$  \\
          &   	       &status     &            &             & depleted  \\
\hline
Inactive  & 4077       & 3205      & 1043       & 2162        & 682\\
          &            & 78.6\,\%  &  25.6\,\%  & 53.0\,\%    &  16.7\,\%\\
          &            & (100\,\%) & (32.6\,\%) & (67.4\,\%)  & \\
Active    & 340        & 200       & 55         & 145         &  34\\
(7.7\,\%) &            & 58.8\,\%  & 16.2\,\%   & 42.7\,\%    & 10.0\,\%\\
          &            & (100\,\%) & (27.5\,\%) & (72.5\,\%)  & \\
Frac. (\%) &           &           &  6.6       & 8.2         & \\         
   
\hline
          & $0.5<M<1.0$ &      &         &          &  \\
\hline
Inactive  & 447         & 365       & 27         & 338        & 28\\
          &             & 81.7\,\%  & 6.0\,\%    & 75.6\,\%   & 6.3\,\%\\
          &             & (100\,\%) & (7.4\,\%)  & (92.6\,\%) & \\
Active    & 70          & 33        & 1          & 32         & 2\\
(13.5\,\%)&             & 47.1\,\%  & 1.4\,\%    & 45.7\,\%   & 2.9\,\%\\
          &             & (100\,\%) & (3.0\,\%)  & (97.0\,\%) & \\
Frac. (\%) &            &           &  6.0       & 14.1       & \\         
\hline
          & $1.0<M<1.5$ &           &            &            &   \\
\hline
Inactive  &  2280       &  1834     & 765        & 1069       & 275 \\
          &             & 80.4\,\%  & 33.6\,\%   & 46.9\,\%   & 12.1\,\%\\
          &             & (100\,\%) & (41.7\,\%) & (58.3\,\%) & \\
Active    &  73         &  38       &  15        & 23         & 8\\
(3.1\,\%) &             & 52.1\,\%  & 20.5\,\%   & 31.5\,\%   & 11.0\,\%\\
          &             & (100\,\%) & (39.5\,\%) & (60.5\,\%) & \\
Frac. (\%) &            &           &  2.9       & 3.2        & \\         
\hline
          & $1.5<M<2.0$ &           &            &            &     \\
\hline
Inactive  &   1046      &   737     &  235       &   502      & 323\\
          &             & 70.5\,\%  & 22.5\,\%   & 48.0\,\%   & 30.9\,\%\\
          &             & (100\,\%) & (31.9\,\%) & (68.1\,\%) & \\
Active    &    68       &  46       &  21        & 25         & 5\\
(6.1\,\%) &             & 67.6\,\%  & 30.9\,\%   & 36.8\,\%   & 7.4\,\%\\
          &             & (100\,\%) & (45.7\,\%) & (54.3\,\%) & \\
Frac. (\%) &            &           &  8.5       & 4.9        & \\         
\hline
          & $2.0<M<3.0$ &           &            &            &      \\
\hline
Inactive  &   311       &  268      &   16      &   252      & 61\\
          &             & 86.2\,\%  & 5.1\,\%   & 81.0\,\%   & 19.6\,\%\\
          &             & (100\,\%) & (6.0\,\%) & (94.0\,\%) & \\
Active    &    125      &  80       &   16      &  64        & 18\\
(28.7\,\%)&             & 64.0\,\%  & 12.8\,\%  & 51.2\,\%   & 14.4\,\%\\
          &             & (100\,\%) & (20.0\,\%)& (80.0\,\%) & \\
Frac. (\%) &            &           &  57.4     & 25.5       & \\ 
\hline
\multicolumn{6}{c}{Low-mass range used in \citet{Tayar_2015} and \citet{Ceillier_2017} }  \\
\multicolumn{6}{c}{$0.5<M<1.1$}  \\
\hline
Inactive  & 808         & 680       & 102        & 578        & 54\\
          &             & 84.2\,\%  & 12.6\,\%   & 71.5\,\%   & 6.7\,\%\\
          &             & (100\,\%) & (15.0\,\%) & (85.0\,\%) & \\
Active    & 95          & 45        & 2          & 43         & 4\\
(10.5\,\%)&             & 47.4\,\%  & 2.1\,\%    & 45.3\,\%   & 4.2\,\%\\
          &             & (100\,\%) & (4.4\,\%)  & (95.6\,\%) & \\
Frac. (\%) &            &           &  3.4       & 11.7       & \\      
\hline
\end{tabular}
\end{table}

\section{Search for contaminants}
\label{app_contaminants}
We have investigated the presence of nearby stars with the targets we monitored with the 3.5-m telescope at APO to flag possible false positive active RGs. We downloaded the target pixel files from the Kepler MAST and plotted the light curve of each individual pixel for each target to see if the rotational modulation was leaking from a nearby star (see details in \citealt{Gaulme_2013}).

Most of them look fine, being the only target in the field of view. However, we note possible contamination for the following. There are two cases of two stars with similar magnitudes that are nearby in the field of view, but still separable: KIC 3458643 (10.2 and 10.8 magnitude possible RGs), and KIC 12266731.

Besides, three targets are inseparable from a nearby star (1-2 pixel difference in position), but the apparent companions are faint enough to not likely matter. GAIA also lists the brighter star in both cases as a giant. These targets are KIC 5166899, where a 10.8 mag RG overlaps with a 13.8 magnitude nearby star, KIC 8326469 (8.3 mag RG and 12.7 magnitude apparent companion), and KIC 11037219 (10.2 mag RG and 14.1 magnitude apparent companion). Another case of stars that lie within 1 or 2 pixels is KIC 5112741:  it is composed of a 12.4 and a 12.6 mag stars. However, according to the pixel files, the activity signal appears to come from the target. 

The most suspicious targets with possible contaminants are: KIC 2833697 (and nearby KIC 2833701) with 9.3 and 9.5 magnitudes respectively, both listed as RGs by GAIA data, and KIC 8160175 with a 10.9 and a 12.2 magnitude star. In the latter case, the photometric signal seem to come from the fainter star. These observations about possible contamination are reported in the ``comment'' column of Table \ref{table_ze_table_p1}.



\clearpage
\newpage

\begin{sidewaystable}
\vspace{8cm}
\tiny
\caption{\tiny Properties of the targets observed with the ARCES spectrometer. Columns report KIC identification number, number of epochs of spectroscopic observations, GAIA distance $d\ind{gaia}$ and radius $R\ind{gaia}$, asteroseismic radius $R\ind{ast}$, and mass $M\ind{ast}$, GAIA effective temperature $T\ind{eff,gaia}$, SDSS/APOGEE (DR16) temperature $T\ind{eff,apg}$, surface gravity $\log g\ind{apg}$ and metallicity $[Fe/H]\ind{apg}$, height of oscillation envelope $H\ind{osc}$, frequency at maximum amplitude $\nu\ind{max}$ and mean frequency separation $\Delta\nu$ of the oscillations, mean $l=1$ mixed-mode period spacing $\Delta\Pi_1$, photometric activity index $S\ind{ph}$, rotation period $P\ind{rot}$, RV standard deviation $\sigma\ind{RV}$, and $v\sin(i=90^\circ)$. Evolutionary status, comments, and literature information are reported in the last three columns. When errors are available, the least significant digits in brackets after the value indicates the uncertainty. Among the comments, the abbreviation ``BF $x$ km s$^{-1}$'' indicates the width of the broadening function \citep{Rucinski_2002}, which is used to point out fast rotators. The abbreviation ``Li line'' or ``Li rich'' indicate when the Lithium absorption line at 6707.7 \AA\ is clearly detectable or very deep. Literature abbreviation: \citetalias{Balona_2015} \citet{Balona_2015}; \citetalias{Ceillier_2017} \citet{Ceillier_2017}; \citetalias{Costa_2015} \citet{Costa_2015}; \citetalias{Davenport_2016} \citet{Davenport_2016}; \citetalias{Deacon_2016} \citet{Deacon_2016}; \citetalias{Godoy-Rivera_2018} \citet{Godoy-Rivera_2018}; \citetalias{Howell_2016} \citet{Howell_2016}; \citetalias{Honda_2016} \citet{Honda_2016}; \citetalias{Katsova_2018} \citet{Katsova_2018}; \citetalias{Kumar_2018} \citet{Kumar_2018}; \citetalias{Pugh_2016} \citet{Pugh_2016}; \citetalias{Schwamb_2012} \citet{Schwamb_2012}; \citetalias{Tang_2012} \citet{Tang_2012}; \citetalias{Tayar_2015} \citet{Tayar_2015}
}
\label{table_ze_table_p1}
\centering
\scalebox{0.7}{
\begin{tabular}
{ p{.04\textwidth}  p{.01\textwidth} p{.03\textwidth} p{.05\textwidth}  p{.05\textwidth} p{.05\textwidth}  p{.05\textwidth} p{.05\textwidth} p{.05\textwidth}  p{.05\textwidth} p{.04\textwidth}  p{.04\textwidth}  p{.04\textwidth}  p{.04\textwidth}  p{.03\textwidth} p{.04\textwidth} p{.035\textwidth} p{.035\textwidth} p{.035\textwidth} p{.035\textwidth} p{.12\textwidth} p{.14\textwidth}} 
\hline 
KIC & 	Visit & $m\ind{Kep}$ & $d\ind{gaia}$ & $R\ind{gaia}$ & $R\ind{ast}$ & $M\ind{ast}$ & $T\ind{eff,gaia}$ & $T\ind{eff,apg}$ & $\log g\ind{apg}$ & $[Fe/H]\ind{apg}$ & $H\ind{osc}$  & $\nu\ind{max}$ & $\Delta\nu$ & $\Delta\Pi_1$	& 	$S\ind{ph}$	& $P\ind{rot}$	& $\sigma\ind{RV}$ & $v\sin 90$& Evol & Comment	&	Literature\\
    & 	    	&  & pc & $R_\odot$ & $R_\odot$ & $M_\odot$ & K  & K & dex & dex & ppm$^2 \mu$Hz$^{-1}$  & $\mu$Hz & $\mu$Hz & s	& \% &  days	& km s$^{-1}$ &	km s$^{-1}$ & 	& 	&\\
\hline 
\multicolumn{22}{c}{3.5m APO without oscillations} \\ 
\hline 
2441154 & 1 & 10.33 & 577(8) & 8.91(64) & ... & ... & 4536(159) & ... & ... & 0.00 & ... & ... & ... & ... & 1.32 & 32.0 & 0.00 & ... & ... & SB from APOGEE &  \\ 
2852961 & 4 & 10.15 & 762(15) & 10.05(46) & ... & ... & 4645(93) & ... & ... & ... & ... & ... & ... & ... & 1.10 & 35.5 & 10.41 & ... & ... & SB1, flares, DR16 (2 vis) & Flares \citetalias{Balona_2015, Pugh_2016, Katsova_2018} \\ 
3459199 & 4 & 10.05 & 990(30) & 12.46(64) & ... & ... & 5340(107) & 5008 & 2.64 & -0.09 & ... & ... & ... & ... & 0.00 & 11.6 & 11.83 & 12.0 & ... & SB1, EB signal & EB contamination \citetalias{Schwamb_2012} \\ 
4278024 & 3 & 11.48 & 959(24) & 8.39(63) & ... & ... & 4711(165) & ... & ... & 0.00 & ... & ... & ... & ... & 2.87 & 37.0 & 17.68 & ... & ... & SB1, DR16 (4 vis) &  \\ 
5112741 & 1 & 12.45 & 2711(151) & 15(1) & ... & ... & 4659(93) & ... & ... & 0.00 & ... & ... & ... & ... & 0.63 & 8.8 & 0.00 & 93.2 & ... & SB2 &  \\ 
5166899 & 4 & 10.91 & 876(19) & 8.54(63) & ... & ... & 4762(167) & 4741 & 2.95 & -0.05 & ... & ... & ... & ... & 1.57 & 23.8 & 28.50 & ... & ... & SB2, faint star within 2 pix, faint Li line &  \\ 
5182131 & 2 & 12.41 & 1495(50) & 7.14(56) & ... & ... & 4975(174) & ... & ... & 0.00 & ... & ... & ... & ... & 2.36 & 76.0 & 12.85 & ... & ... & SB1 &  \\ 
5437763 & 2 & 11.79 & 1438(50) & 10.26(81) & ... & ... & 4696(164) & 4523 & 3.04 & -0.15 & ... & ... & ... & ... & 0.56 & 14.0 & 16.72 & ... & ... & SB1. &  \\ 
5630212 & 2 & 10.21 & 560(10) & 9.74(70) & ... & ... & 3999(140) & ... & ... & 0.00 & ... & ... & ... & ... & 3.89 & 31.1 & 15.17 & 3.0 & ... & SB1, faint Li line &  \\ 
5821762 & 3 & 12.30 & 2734(204) & 13(1) & ... & ... & 5209(182) & 5205 & 2.68 & 0.05 & ... & ... & ... & ... & 0.47 & 9.2 & 10.08 & ... & ... & Unclear, BF 120 km s$^{-1}$ &  \\ 
6372268 & 5 & 11.45 & 645(11) & 6.57(48) & ... & ... & 4014(140) & ... & ... & 0.00 & ... & ... & ... & ... & 7.81 & 5.2 & 39.12 & 52.4 & ... & SB2, Kep EB, BF 120 km s$^{-1}$, X-ray & \citetalias{Howell_2016} \\ 
6522973 & 2 & 11.80 & 1877(80) & 15(1) & ... & ... & 4556(159) & 4385 & 2.68 & -0.04 & ... & ... & ... & ... & 0.59 & 30.3 & 2.02 & ... & ... & SB1 &  \\ 
6707805 & 1 & 12.45 & 1621(71) & 6.79(41) & ... & ... & 5259(105) & ... & ... & 0.00 & ... & ... & ... & ... & 1.76 & 22.0 & 0.00 & ... & ... & SB2, faint Li line &  \\ 
7552750 & 2 & 12.29 & 1252(62) & 5.47(48) & ... & ... & 5236(183) & ... & ... & 0.00 & ... & ... & ... & ... & 0.16 & 101.0 & 5.20 & ... & ... & SB1 &  \\ 
7869590 & 7 & 10.88 & 629(10) & 7.18(52) & ... & ... & 4647(163) & 4370 & 3.22 & 0.26 & ... & ... & ... & ... & 3.53 & 40.5 & 13.84 & 9.5 & ... & SB1, no clear H$_\alpha$ line & Long term var \citetalias{Tang_2012} \\ 
7899428 & 1 & 11.34 & 598(10) & 4.53(33) & ... & ... & 4915(172) & ... & ... & 0.00 & ... & ... & ... & ... & 3.71 & 11.5 & 0.00 & ... & ... & SB3, Li line &  \\ 
8517303 & 2 & 11.43 & 851(18) & 8.06(38) & ... & ... & 4554(91) & 4547 & 3.08 & -0.10 & ... & ... & ... & ... & 5.51 & 53.5 & 6.41 & ... & ... & SB1 &  \\ 
8749284 & 3 & 12.19 & 1002(25) & 7.97(38) & ... & ... & 4424(88) & ... & ... & ... & ... & ... & ... & ... & 1.52 & 3.5 & 8.34 & ... & ... & Unclear, BF 100 km s$^{-1}$, H$_\alpha$ em. &  \\ 
8822505 & 3 & 11.80 & 1369(48) & 10.30(81) & ... & ... & 4729(166) & 4422 & 2.91 & -0.09 & ... & ... & ... & ... & 3.01 & 40.0 & 7.90 & ... & ... & SB1 &  \\ 
9770992 & 4 & 10.53 & 1249(98) & 14(2) & ... & ... & 4921(172) & ... & ... & 0.00 & ... & ... & ... & ... & 1.78 & 17.8 & 35.93 & ... & ... & SB2 &  \\ 
10297952 & 2 & 11.49 & 1215(34) & 9.29(71) & ... & ... & 3787(133) & 4484 & 3.33 & -0.51 & ... & ... & ... & ... & 3.01 & 17.6 & 10.16 & ... & ... & SB1, DR16 (2 vis) &  \\ 
10908111 & 2 & 12.12 & 1585(65) & 8.83(72) & ... & ... & 4908(172) & 4473 & 2.55 & -0.13 & ... & ... & ... & ... & 1.54 & 13.6 & 29.04 & 3.1 & ... & SB1/2 &  \\ 
10991377 & 3 & 12.12 & 1950(112) & 12(1) & ... & ... & 4872(171) & 4587 & 2.63 & -0.07 & ... & ... & ... & ... & 1.36 & 15.7 & 7.00 & 5.0 & ... & Unclear, BF 75 km s$^{-1}$, Li rich &  \\ 
11146520 & 2 & 12.42 & 1518(51) & 7.78(61) & ... & ... & 4832(169) & 4608 & 2.87 & -0.29 & ... & ... & ... & ... & 1.64 & 34.0 & 9.39 & ... & ... & SB1 &  \\ 
11551404 & 4 & 0.00 & 325(2) & 2.67(19) & ... & ... & 4968(174) & 4703 & 3.62 & -0.16 & ... & ... & ... & ... & 0.00 &  & 14.26 &  & ... & SB2, weak H$_\alpha$ abs., faint Li line & Flares, wide binary candidate \citetalias{Davenport_2016,Deacon_2016,Godoy-Rivera_2018} \\ 
11554998 & 4 & 11.78 & 1082(26) & 8.16(61) & ... & ... & 4727(165) & 4301 & 3.03 & -0.13 & ... & ... & ... & ... & 5.80 & 30.4 & 22.04 & ... & ... & SB1 &  \\ 
11907579 & 3 & 12.06 & 2105(101) & 12(1) & ... & ... & 4998(175) & 4679 & 2.84 & 0.08 & ... & ... & ... & ... & 0.53 & 14.9 & 0.27 & ... & ... & RV stable, BF 60 km s$^{-1}$, Li line &  \\ 
12118682 & 2 & 11.89 & 1674(66) & 10.77(87) & ... & ... & 4918(172) & 4515 & 2.81 & -0.26 & ... & ... & ... & ... & 2.03 & 45.0 & 18.24 & ... & ... & SB1 &  \\ 
12157366 & 2 & 12.47 & 1765(87) & 11.21(97) & ... & ... & 4483(157) & ... & ... & 0.00 & ... & ... & ... & ... & 1.67 & 18.1 & 2.77 & 3.9 & ... & SB1 &  \\ 
12406908 & 1 & 12.35 & 789(13) & 4.47(32) & ... & ... & 4585(160) & ... & ... & 0.00 & ... & ... & ... & ... & 5.30 & 13.4 & 0.00 & ... & ... & SB2, H$_\alpha$ em., Ca II H \ & K, triplet \\ 
\hline 
\multicolumn{22}{c}{3.5m APO with oscillations} \\ 
\hline 
3654361 & 5 & 11.27 & 732(15) & 4.77(22) & 4.24(44) & 1.39(43) & 5066(101) & 5109 & 3.36 & -0.22 & ... & 257(26) & 18.10 & 82.70 & 0.25 & 41.5 & 0.61 & 5.2 & RGB &  &  \\ 
5430224 & 7 & 11.01 & 1171(30) & 10.81(81) & 11.22(45) & 2.05(25) & 4949(173) & 4899 & 3.19 & 0.05 & 111(19) & 55(2) & 5.09 & ... & 1.11 & 55.0 & 0.70 & 10.3 & RC &  &  \\ 
5514251 & 7 & 8.98 & 451(5) & 10.17(72) & 9.83(31) & 0.884(87) & 4944(173) & ... & ... & 0.00 & 1373(172) & 30.73(78) & 4.08 & 260.90 & 0.44 & 33.0 & 0.85 & 15.1 & RC & Li rich &  \\ 
5559520 & 5 & 10.16 & 766(15) & 10.71(79) & 10.73(27) & 1.078(86) & 4875(171) & ... & ... & 0.00 & 2853(202) & 31.63(54) & 3.95 & 239.80 & 0.25 & 41.0 & 0.74 & 13.2 & RC &  &  \\ 
5633342 & 5 & 9.79 & 497(36) & 6.69(68) & 4.537(85) & 1.58(10) & 5136(180) & ... & ... & 0.00 & 4.74(58) & 253(2) & 17.39 & 94.90 & 0.31 & 109.0 & 0.76 & 2.1 & RGB & DR16 (4 vis) &  \\ 
6221548 & 7 & 10.98 & 1075(42) & 8.62(49) & 8.42(12) & 2.44(11) & 5043(101) & 5104 & 2.82 & -0.06 & 57(3) & 114(1) & 8.55 & 184.10 & 0.82 & 102.0 & 0.83 & 4.2 & RC &  &  \\ 
7103951 & 5 & 9.38 & 515(10) & 9.40(42) & 10.71(16) & 1.795(87) & 4906(98) & 4902 & 2.95 & 0.02 & 637(31) & 52.70(57) & 5.11 & 304.10 & 0.25 & 75.0 & 2.60 & 7.2 & RC & SB1, DR16 (2 vis) & Rapid rot. RG \citetalias{Costa_2015} \\ 
7183551 & 7 & 8.23 & 327(3) & 9.93(70) & 9.99(39) & 1.00(12) & 4953(173) & ... & ... & 0.00 & 860(121) & 34(1) & 4.23 & ... & 0.26 & 36.3 & 0.55 & 13.9 & ... & $\ell=1$ dep. &  \\ 
7451465 & 5 & 11.33 & 1204(36) & 8.36(64) & 8.70(25) & 3.04(28) & 5204(182) & ... & ... & 0.00 & 8(2) & 131(3) & 9.08 & ... & 0.87 & 47.1 & 0.35 & 9.3 & RGB? & DR16 (4 vis) &  \\ 
7465993 & 5 & 10.70 & 748(19) & 7.16(34) & 7.078(92) & 2.148(91) & 5019(100) & 5044 & 2.84 & 0.12 & 35(2) & 143(1) & 10.41 & 54.50 & 0.26 & 81.5 & 0.73 & 4.4 & RGB & $\ell=1$ dep. &  \\ 
7703954 & 5 & 9.42 & 593(8) & 9.95(43) & 9.94(16) & 2.83(15) & 5045(101) & 5099 & 2.81 & 0.07 & 75(4) & 95(1) & 7.18 & 216.00 & 0.30 & 72.0 & 0.71 & 7.0 & RC &  &  \\ 
7743728 & 5 & 10.93 & 957(32) & 6.92(36) & 6.25(17) & 2.41(20) & 5308(106) & 5296 & 3.46 & 0.04 & 7(1) & 200(5) & 13.28 & ... & 0.50 & 38.6 & 0.41 & 8.2 & RGB &  &  \\ 
8160175 & 6 & 10.99 & 782(13) & 7.30(32) & 7.382(79) & 1.270(46) & 4741(95) & 4743 & 2.83 & 0.03 & 840(24) & 79.85(22) & 7.51 & 73.00 & 3.86 & 69.8 & 0.99 & 5.3 & RGB & Likely contamination activity &  \\ 
8326469 & 5 & 8.40 & 334(3) & 8.56(61) & 8.36(17) & 2.44(17) & 5203(182) & ... & ... & 0.00 & 63(3) & 114(1) & 8.64 & 248.30 & 0.48 & 144.0 & 0.66 & 2.9 & RC & faint star within 2 pix & Flares \citetalias{Honda_2016} \\ 
8515227 & 7 & 11.18 & 1172(57) & 10.94(71) & 4.286(74) & 0.939(48) & 4706(94) & 4778 & 3.17 & -0.68 & 11.21(86) & 176(2) & 14.60 & 82.20 & 0.99 & 28.0 & 23.66 & 7.7 & RGB & SB3, DR16 (2 vis) &  \\ 
8816525 & 6 & 8.91 & 499(7) & 10.16(44) & 9.90(15) & 2.28(11) & 5058(101) & 5084 & 2.76 & 0.05 & 215(9) & 77.29(85) & 6.49 & 274.40 & 0.23 & 68.5 & 1.05 & 7.3 & RC & faint Li line &  \\ 
9788756 & 6 & 11.38 & 999(23) & 7.75(58) & 7.94(39) & 1.75(23) & 4843(170) & 4903 & 3.05 & -0.10 & 59(5) & 94(3) & 7.90 & 218.00 & 2.75 & 95.0 & 0.57 & 4.2 & RC & Very low modes &  \\ 
10198347 & 5 & 10.29 & 1020(27) & 11.40(86) & 10.81(24) & 1.42(10) & 5191(182) & 4924 & 2.96 & -0.16 & 2282(139) & 39.65(50) & 4.47 & 276.60 & 0.28 & 45.6 & 2.84 & 12.0 & RC & SB2 & Anom. rot. \citetalias{Tayar_2015} \\ 
10461323 & 7 & 10.01 & 642(9) & 8.56(38) & 8.95(17) & 2.14(13) & 4902(98) & 4874 & 2.99 & 0.12 & 88(5) & 90(1) & 7.30 & ... & 0.53 & 79.7 & 0.66 & 5.7 & ... & $\ell=1$ dep. &  \\ 
10793890 & 6 & 9.95 & 675(13) & 8.04(59) & 8.49(17) & 2.49(16) & 5310(186) & 5113 & 2.79 & 0.04 & 70(3) & 111.51(98) & 8.51 & 267.70 & 0.29 & 130.0 & 0.64 & 3.3 & RC &  &  \\ 
11037219 & 7 & 10.22 & 680(15) & 7.97(59) & 8.49(17) & 2.44(16) & 5064(177) & 4947 & 3.10 & 0.25 & 80(3) & 112.14(98) & 8.44 & 146.50 & 0.34 & 140.0 & 0.75 & 3.1 & RC & faint star within 2 pix &  \\ 
11087027 & 8 & 10.88 & 1238(103) & 14(2) & 13(5) & 2(2) & 4918(172) & ... & ... & ... & 252(94) & 33(12) & 3.70 & ... & 0.63 & 30.5 & 0.77 & 21.1 & ... & Flares, Li rich, $H_\alpha$ em. &  \\ 
12266731 & 5 & 10.21 & 899(19) & 9.72(44) & 9.62(16) & 2.72(14) & 5152(103) & 5104 & 2.84 & 0.15 & 92(5) & 97(1) & 7.40 & 246.90 & 0.21 & 79.0 & 0.56 & 6.2 & RC & Star nearby &  \\ 
\hline																				
\end{tabular}
}
\end{sidewaystable}

\clearpage
\newpage

\begin{sidewaystable}
\vspace{9cm}
\tiny
\caption{\tiny Properties of the targets observed with the APOGEE spectrometer. Some overlap with the targets observed with ARCES, and are indicated in the comment column as ``also 3.5-m''.}
\label{table_ze_table_p2}
\centering
\scalebox{0.7}{
\begin{tabular}
{ p{.04\textwidth}  p{.01\textwidth} p{.03\textwidth} p{.05\textwidth}  p{.05\textwidth} p{.05\textwidth}  p{.05\textwidth} p{.05\textwidth} p{.05\textwidth}  p{.05\textwidth} p{.04\textwidth}  p{.04\textwidth}  p{.04\textwidth}  p{.04\textwidth}  p{.03\textwidth} p{.04\textwidth} p{.035\textwidth} p{.035\textwidth} p{.035\textwidth} p{.035\textwidth} p{.12\textwidth} p{.14\textwidth}} 
\hline 
KIC & 	Visit & $m\ind{Kep}$ & $d\ind{gaia}$ & $R\ind{gaia}$ & $R\ind{ast}$ & $M\ind{ast}$ & $T\ind{eff,gaia}$ & $T\ind{eff,apg}$ & $\log g\ind{apg}$ & $[Fe/H]\ind{apg}$ & $H\ind{osc}$  & $\nu\ind{max}$ & $\Delta\nu$ & $\Delta\Pi_1$	& 	$S\ind{ph}$	& $P\ind{rot}$	& $\sigma\ind{RV}$ & $v\sin 90$& Evol & Comment	&	Literature\\
    & 	    	&  & pc & $R_\odot$ & $R_\odot$ & $M_\odot$ & K  & K & dex & dex & ppm$^2 \mu$Hz$^{-1}$  & $\mu$Hz & $\mu$Hz & s	& \% &  days	& km s$^{-1}$ &	km s$^{-1}$ & 	& 	&\\
\hline 
\multicolumn{22}{c}{APOGEE with more than three visits} \\ 
\hline 
2305930 & 3 & 10.80 & 940(21) & 9.77(72) & 9.48(22) & 0.760(58) & 4891(171) & 4859 & 2.91 & -0.43 & 5151(303) & 28.49(42) & 3.99 & 247.70 & 0.21 & 33.0 & 0.11 & 14.5 & RC &  & Li rich \citetalias{Kumar_2018}, rapid rot. \citetalias{Tayar_2015} \\ 
2441154 & 5 & 10.33 & 577(8) & 8.91(64) & ... & ... & 4536(159) & ... & ... & 0.00 & ... & ... & ... & ... & 1.32 & 32.0 & 1.65 & ... & ... & SB (also 3.5-m) &  \\ 
2833697 & 3 & 9.42 & 565(9) & 9.47(41) & 9.35(14) & 0.789(38) & 5001(100) & 5028 & 2.44 & -0.81 & 5710(265) & 30.13(32) & 4.16 & 345.20 & 0.11 & 41.5 & 0.09 & 11.4 & RC & Possible contamination from KIC 2833701 & Contamination? \citetalias{Ceillier_2017} \\ 
3216467 & 3 & 11.21 & 1088(31) & 10.71(81) & 11.04(23) & 1.108(76) & 4646(163) & 4508 & 2.48 & 0.43 & 6990(292) & 31.47(32) & 3.84 & ... & 0.24 & 163.0 & 0.06 & 3.4 & ... &  &  \\ 
3339894 & 3 & 11.34 & 1586(69) & 13(1) & 13.96(30) & 1.74(12) & 4779(167) & 4749 & 2.44 & -0.22 & 2633(145) & 30.47(33) & 3.38 & ... & 0.14 & 100.0 & 0.16 & 7.1 & ... &  &  \\ 
3432732 & 3 & 10.74 & 1162(33) & 11.01(84) & 10.38(22) & 2.59(18) & 5123(179) & 4986 & 2.67 & 0.21 & 168(8) & 79.44(93) & 6.44 & 248.00 & 0.12 & 117.0 & 0.09 & 4.5 & RC & $\ell=1$ dep. & 122-d rot \citetalias{Ceillier_2017} \\ 
3458643 & 3 & 10.23 & 888(22) & 10.39(50) & 11.05(29) & 2.55(21) & 5008(100) & 5035 & 2.74 & 0.08 & 121(11) & 70(2) & 5.82 & 213.90 & 0.54 & 54.0 & 2.19 & 10.3 & RC & SB, star nearby &  \\ 
3660182 & 4 & 12.06 & 1895(115) & 12(1) & 9.08 & 0.69 & 4784(167) & ... & ... & 0.00 & 5132(279) & 28.59(33) & 4.06 & ... & 0.00 &  & 0.08 &  & ... &  &  \\ 
3837107 & 3 & 10.19 & 650(17) & 7.83(38) & 8.23(10) & 2.111(85) & 4979(100) & 4933 & 2.69 & 0.13 & 123(4) & 104.12(69) & 8.22 & ... & 0.13 & 93.0 & 0.03 & 4.5 & RGB? &  &  \\ 
3937217 & 3 & 11.63 & 1511(54) & 11.03(87) & 10.78(25) & 0.971(73) & 4853(170) & 4759 & 2.52 & -0.08 & 5759(335) & 28.29(41) & 3.72 & 278.90 & 0.18 & 57.0 & 0.03 & 9.6 & RC &  & Surf. rot. \citetalias{Costa_2015, Ceillier_2017} \\ 
3967501 & 4 & 9.93 & 408(4) & 6.26(45) & 6.06(42) & 1.23(26) & 4769(167) & 4682 & 3.09 & 0.29 & 46(12) & 115(8) & 9.95 & 146.90 & 1.49 & 85.5 & 3.40 & 3.6 & RC & SB & Flares \citetalias{Honda_2016} \\ 
4242873 & 5 & 12.49 & 1317(35) & 5.94(29) & 5.758(72) & 1.190(49) & 4858(97) & 4848 & 3.12 & -0.02 & 74(4) & 121.48(89) & 10.56 & 63.90 & 0.28 & 74.0 & 4.05 & 3.9 & RGB & SB &  \\ 
4278024 & 4 & 11.48 & 959(24) & 8.39(63) & ... & ... & 4711(165) & ... & ... & 0.00 & ... & ... & ... & ... & 2.87 & 37.0 & 8.59 & ... & ... & SB (also 3.5-m) &  \\ 
4348593 & 3 & 8.97 & 620(12) & 12.72(93) & 12.45(26) & 2.94(21) & 5065(177) & 5071 & 2.73 & 0.11 & 264(13) & 62.96(72) & 5.22 & 273.20 & 0.18 & 107.0 & 0.04 & 5.9 & RC &  &  \\ 
4659808 & 4 & 11.31 & 1543(65) & 10.97(90) & 10.93(26) & 2.70(21) & 5166(181) & 5062 & 2.73 & 0.07 & 138(8) & 74(1) & 6.08 & 228.70 & 0.56 & 75.0 & 0.11 & 7.4 & RC &  &  \\ 
5633342 & 4 & 9.79 & 497(36) & 6.69(68) & 4.537(85) & 1.58(10) & 5136(180) & ... & ... & 0.00 & 4.74(58) & 253(2) & 17.39 & 94.90 & 0.31 & 109.0 & 0.05 & 2.1 & RGB & (also 3.5-m) &  \\ 
6207670 & 4 & 11.90 & 1712(59) & 11.47(90) & 10.42(22) & 1.009(69) & 4713(165) & ... & ... & 0.00 & 5927(300) & 31.96(31) & 4.00 & ... & 0.23 & 121.0 & 0.03 & 4.4 & ... &  &  \\ 
6447237 & 4 & 12.43 & 1638(63) & 6.75(54) & 6.33(18) & 2.02(18) & 5152(180) & ... & ... & 0.00 & 15(3) & 166(4) & 11.94 & ... & 0.09 & 140.0 & 0.37 & 2.3 & RGB & Not an SB &  \\ 
7351098 & 3 & 11.20 & 868(29) & 6.60(52) & 7.98(15) & 2.22(15) & 5057(177) & 4916 & 3.08 & 0.32 & 97(4) & 115.82(92) & 8.84 & 148.70 & 0.26 & 125.0 & 0.05 & 3.2 & RC &  &  \\ 
7451465 & 4 & 11.33 & 1204(36) & 8.36(64) & 8.70(25) & 3.04(28) & 5204(182) & ... & ... & 0.00 & 8(2) & 131(3) & 9.08 & ... & 0.87 & 47.1 & 0.11 & 9.3 & RGB? & (also 3.5-m) &  \\ 
7670419 & 3 & 11.31 & 1299(50) & 11.75(94) & 11.15(27) & 1.086(84) & 4632(162) & 4558 & 2.60 & 0.28 & 4460(263) & 30.30(48) & 3.74 & 236.50 & 0.43 & 108.0 & 0.03 & 5.2 & RC &  & Rapid rot. \citetalias{Costa_2015} \\ 
7671562 & 3 & 11.65 & 1439(43) & 10.41(79) & 10.83(22) & 1.027(69) & 4721(165) & 4619 & 2.43 & 0.11 & 7554(349) & 30.09(27) & 3.80 & ... & 0.27 & 114.0 & 0.05 & 4.8 & ... &  &  \\ 
7948193 & 3 & 11.09 & 896(19) & 6.89(50) & 7.53(15) & 1.97(13) & 5216(183) & 4921 & 3.10 & 0.05 & 85(3) & 113.70(87) & 9.09 & ... & 0.37 & 109.0 & 1.01 & 3.5 & RGB? & SB, $\ell=1$ dep. &  \\ 
8013399 & 3 & 11.58 & 735(12) & 4.32(19) & 2.465(46) & 0.390(23) & 5028(101) & 5100 & 3.47 & 0.00 & 7(2) & 213(3) & 21.56 & ... & 0.10 & 120.0 & 0.14 & 1.0 & RGB & Contamination? &  \\ 
8085964 & 3 & 9.89 & 718(11) & 9.47(68) & 9.08(19) & 2.37(16) & 5099(178) & 4981 & 2.79 & 0.22 & 162(8) & 95.16(100) & 7.53 & 212.50 & 0.16 & 97.0 & 0.08 & 4.7 & RC &  & Rapid rot. \citetalias{Costa_2015} \\ 
8479301 & 5 & 12.22 & 1252(41) & 5.83(30) & 5.926(73) & 1.371(56) & 4836(97) & 4862 & 3.11 & -0.33 & 57(3) & 132.42(92) & 10.85 & 71.80 & 0.49 & 59.1 & 0.15 & 5.1 & RGB & $\ell=1$ dep. &  \\ 
8547469 & 5 & 11.85 & 1742(61) & 10.48(82) & 10.64(23) & 0.992(71) & 4936(173) & 4694 & 2.41 & -0.06 & 5914(311) & 29.39(36) & 3.83 & ... & 0.21 & 80.5 & 0.13 & 6.7 & ... &  &  \\ 
8936339 & 5 & 12.11 & 2014(83) & 10.47(86) & 11.17(25) & 1.65(12) & 5070(177) & 4833 & 2.75 & 0.06 & 1135(62) & 43.89(56) & 4.61 & 274.20 & 0.24 & 103.0 & 0.14 & 5.5 & RC &  &  \\ 
8941031 & 5 & 12.08 & 1873(94) & 10.88(71) & 11.22(24) & 1.019(65) & 4659(93) & 4660 & 2.68 & -0.09 & 4714(359) & 27.96(49) & 3.59 & 201.20 & 0.27 & 54.2 & 0.16 & 10.5 & RC & $\ell=1$ dep. &  \\ 
9008125 & 3 & 11.88 & 981(22) & 5.08(24) & 5.30(16) & 1.94(18) & 4975(100) & 4996 & 3.27 & -0.14 & 6(2) & 231(6) & 15.27 & ... & 0.76 & 88.0 & 0.16 & 3.0 & RGB & $\ell=1$ dep. &  \\ 
9222065 & 6 & 9.83 & 468(5) & 5.86(42) & 5.95(23) & 1.81(21) & 5176(181) & 5044 & 2.87 & 0.09 & 27(6) & 168(6) & 12.39 & 85.80 & 0.30 & 88.5 & 0.05 & 3.4 & RGB & $\ell=1$ dep. &  \\ 
9513046 & 3 & 10.56 & 1007(29) & 8.92(45) & 9.21(16) & 2.09(11) & 5145(103) & 5170 & 2.67 & -0.38 & 119(8) & 81(1) & 6.91 & 161.30 & 0.22 & 98.0 & 0.01 & 4.8 & RC & $\ell=1$ dep. &  \\ 
9837673 & 6 & 11.65 & 997(22) & 5.52(41) & 5.277(100) & 1.88(12) & 5177(181) & 5062 & 3.26 & -0.20 & 28(1) & 222(2) & 15.13 & 96.10 & 0.24 & 83.0 & 0.19 & 3.2 & RGB & SB, $\Delta V = 0.5$ km s$^{-1}$ &  \\ 
10118979 & 3 & 11.77 & 1634(54) & 10.60(82) & 10.94(23) & 1.052(72) & 4980(174) & 4677 & 2.42 & -0.02 & 10167(496) & 29.40(29) & 3.79 & ... & 0.17 & 118.0 & 0.02 & 4.7 & ... &  &  \\ 
10319146 & 3 & 11.00 & 1118(28) & 10.65(79) & 10.67(22) & 1.002(67) & 4997(175) & 4624 & 2.37 & -0.02 & 9162(414) & 29.35(26) & 3.84 & ... & 0.18 & 109.0 & 0.04 & 5.0 & ... &  &  \\ 
11175907 & 3 & 11.57 & 1794(70) & 11.56(93) & 11.78(23) & 1.66(11) & 5194(182) & 4862 & 2.53 & -0.01 & 1873(84) & 39.17(31) & 4.26 & 284.10 & 0.27 & 128.0 & 0.03 & 4.7 & RC &  &  \\ 
\hline 
\multicolumn{22}{c}{APOGEE with two visits} \\ 
\hline 
2852961 & 2 & 10.15 & 762(15) & 10.05(46) & ... & ... & 4645(93) & ... & ... & ... & ... & ... & ... & ... & 1.10 & 35.5 & 1.32 & ... & ... & SB1 (also 3.5-m) &  \\ 
5382824 & 2 & 10.41 & 1025(48) & 10.63(69) & 8.98(12) & 2.52(11) & 5035(101) & 5114 & 2.81 & 0.01 & 80(3) & 104.05(81) & 7.90 & 252.20 & 0.15 & 94.0 & 0.98 & 4.8 & RC & SB, $\ell=1$ dep. & SB, surf. rot. \citetalias{Ceillier_2017} \\ 
5439339 & 2 & 11.27 & 1259(31) & 8.81(42) & 8.70(13) & 2.27(11) & 5082(102) & 5098 & 2.77 & 0.06 & 102(5) & 99(1) & 7.86 & 250.20 & 0.23 & 59.0 & 11.75 & 7.5 & RC & SB & SB, surf. rot. \citetalias{Ceillier_2017} \\ 
6032639 & 2 & 11.56 & 1563(54) & 11.27(60) & 10.46(19) & 1.455(84) & 4864(97) & 4862 & 2.58 & 0.02 & 943(54) & 45.00(68) & 4.77 & 289.80 & 0.23 & 130.0 & 0.88 & 4.1 & RC & SB & SB, surf. rot. \citetalias{Ceillier_2017} \\ 
6590195 & 2 & 12.30 & 1259(35) & 6.42(32) & 6.522(75) & 1.412(54) & 4791(96) & 4779 & 3.05 & 0.25 & 197(7) & 113.11(60) & 9.54 & ... & 0.23 & 92.0 & 14.48 & 3.6 & RGB & SB &  \\ 
6933666 & 2 & 12.07 & 1964(79) & 11.61(66) & 11.29(19) & 1.202(61) & 4669(93) & 4668 & 2.57 & -0.22 & 2849(176) & 32.60(39) & 3.87 & ... & 0.23 & 99.0 & 6.34 & 5.8 & ... & SB & SB, surf. rot. \citetalias{Ceillier_2017} \\ 
7103951 & 2 & 9.38 & 515(10) & 9.40(42) & 10.71(16) & 1.795(87) & 4906(98) & 4902 & 2.95 & 0.02 & 637(31) & 52.70(57) & 5.11 & 304.10 & 0.25 & 75.0 & 1.42 & 7.2 & RC & SB (also 3.5-m) &  \\ 
7969754 & 2 & 12.27 & 1047(28) & 6.53(49) & ... & ... & 4832(169) & 4554 & 3.12 & -0.03 & ... & ... & ... & ... & 3.74 & 69.1 & 25.23 & ... & ... & SB, no osc. &  \\ 
8108336 & 2 & 12.05 & 1790(92) & 11.27(99) & 12.16(27) & 2.04(15) & 4868(170) & ... & ... & 0.00 & 632(37) & 46.60(57) & 4.50 & ... & 0.39 & 10.0 & 4.18 & 61.5 & ... & SB &  \\ 
8515227 & 2 & 11.18 & 1172(57) & 10.94(71) & 4.286(74) & 0.939(48) & 4706(94) & 4778 & 3.17 & -0.68 & 11.21(86) & 176(2) & 14.60 & 82.20 & 0.99 & 28.0 & 5.80 & 7.7 & RGB & SB &  \\ 
8720055 & 2 & 10.40 & 1114(32) & 13.49(69) & 27.03 & 12.01 & 5007(100) & 4974 & 3.04 & -0.05 & 441(18) & 54.81(42) & 3.30 & ... & 0.00 &  & 2.44 &  & ... & $\gamma$ Dor companion? & Flares \citetalias{Davenport_2016} \\ 
9240941 & 2 & 11.35 & 945(21) & 8.09(39) & 8.37(12) & 2.27(11) & 5075(102) & 5068 & 2.76 & 0.03 & 80(4) & 107(1) & 8.32 & ... & 0.12 & 123.0 & 0.50 & 3.4 & ... & $\ell=1$ dep. & Likely SB, surf. rot. \citetalias{Ceillier_2017} \\ 
9651996 & 2 & 11.04 & 1193(33) & 12.18(92) & ... & ... & 4750(166) & 4494 & 2.90 & -0.38 & ... & ... & ... & ... & 3.98 & 36.6 & 3.52 & ... & ... & SB, no osc. &  \\ 
10297952 & 2 & 11.49 & 1215(34) & 9.29(71) & ... & ... & 3787(133) & 4484 & 3.33 & -0.51 & ... & ... & ... & ... & 3.01 & 17.6 & 0.97 & ... & ... & SB (also 3.5-m) &  \\ 
10811720 & 2 & 11.87 & 1845(76) & 10.39(85) & 11.17(22) & 1.487(99) & 5198(182) & 4893 & 2.46 & -0.09 & 2091(100) & 38.98(33) & 4.37 & ... & 0.18 & 110.0 & 1.36 & 5.1 & ... & SB &  \\ 
12314910 & 2 & 11.36 & 1420(50) & 12.76(69) & 13.80(22) & 1.300(63) & 4523(90) & 4514 & 2.39 & -0.18 & 7714(403) & 23.96(25) & 2.97 & ... & 0.42 & 104.0 & 2.81 & 6.7 & ... & SB & SB, surf. rot. \citetalias{Ceillier_2017} \\ 
\hline																				
\end{tabular}
}
\end{sidewaystable}

\end{document}